\documentclass{article}[11pt]
\usepackage[large,bf,FIGTOPCAP]{subfigure}
\usepackage{lineno}
\usepackage{setspace}
\usepackage{graphicx}
\usepackage{color}
\usepackage{epsfig}
\usepackage{amsmath, amsfonts} 
\usepackage{amssymb} 
\usepackage{latexsym}
\usepackage[colorlinks,bookmarks,urlcolor=blue,citecolor=blue,linkcolor=blue]{hyperref}
\usepackage{cite}
\usepackage[T1]{fontenc}
\usepackage{bbm}

\newcommand{\qedsymbol}{\vbox{\hrule height0.6pt\hbox{\vrule height1.3ex width0.6pt\hskip0.8ex\vrule width0.6pt}\hrule height0.6pt}}

\newcommand{\dif}{\frac{\sigma^{2}}{2}}

\title{The Dynamics of Bistable Switching Behavior in Limit Cycle Systems with Additive Noise}
\author{Michael A. Schwemmer\footnote{schwemmer.2@mbi.osu.edu}~~and Jay M. Newby\footnote{newby.23@mbi.osu.edu} \\
{\small Mathematical Biosciences Institute, The Ohio State University, 1735 Neil Ave. Columbus, OH 43210}}
\date{}
\numberwithin{equation}{section}
\begin{document}

\maketitle

\singlespacing


\vspace{-1cm}
\begin{abstract}

Additive noise is known to produce counter-intuitive behaviors in nonlinear dynamical systems.  Previously, it was shown that systems with a deterministic limit cycle can display bistable switching between metastable states in the presence of asymmetric additive white noise.  Here, we systematically analyze the dynamics of this bistable behavior and show how the vector field away from the limit cycle influences the rate and directionality of the bistable switching.  Using stochastic phase reduction methods, we identify mechanisms underlying different rates of switching and predict when the system will rotate in the opposite direction of the deterministic limit cycle.  Thus, this work presents an alternative mechanism for generating a range of bistable switch-like behaviors that have been observed in a number of physical systems.

\end{abstract}

\vspace{.5cm}
\noindent {\bf AMS Subject Classification:} 60H10, 34E10, 92B25 \\

\noindent {\bf Key words:} stochastic differential equations, limit cycle oscillator, phase reduction, bistable switching


\section{Introduction}

Countless physical systems display dynamics that are modeled by limit cycle oscillators \cite{kuramoto,syncbook,winfree}.  As many of these systems are subjected to stochastic fluctuations, there has been a growing interest in understanding how noise affects the dynamics of limit cycle oscillators \cite{bolandetal2009,gonzeetal2002,koeppletal2011,teramae2009,yoshimura2008}. 
When the noise is weak, formal reduction methods can be performed to reduce the dimensionality of the system to a so-called phase equation \cite{teramae2009, yoshimura2008}.  In this case, it is known that the magnitude of the added noise shifts the mean frequency of oscillations away from the natural frequency of the limit cycle. On the other hand, when the noise is large, the trajectories of the system bear no resemblance to the deterministic limit cycle behavior.

The case of moderate noise applied to a limit cycle oscillator has received less attention.  In this case, the vector field around the limit cycle has a strong influence on the dynamics of the stochastic system, as the trajectory is constantly being perturbed off the limit cycle.  In previous work \cite{newby2014}, we showed that an oscillator subject to moderate noise can display numerous interesting and counter-intuitive behaviors. In some cases, the addition of noise can act to completely eliminate oscillations, and even cause the trajectories to rotate in the {\it opposite} direction of the deterministic limit cycle.  Perhaps more interesting is the emergence of bistable switching behavior which occurs in noisy non-rotationally symmetric systems. 

\begin{figure}[h!]
\begin{center}
{\includegraphics[scale=.3]{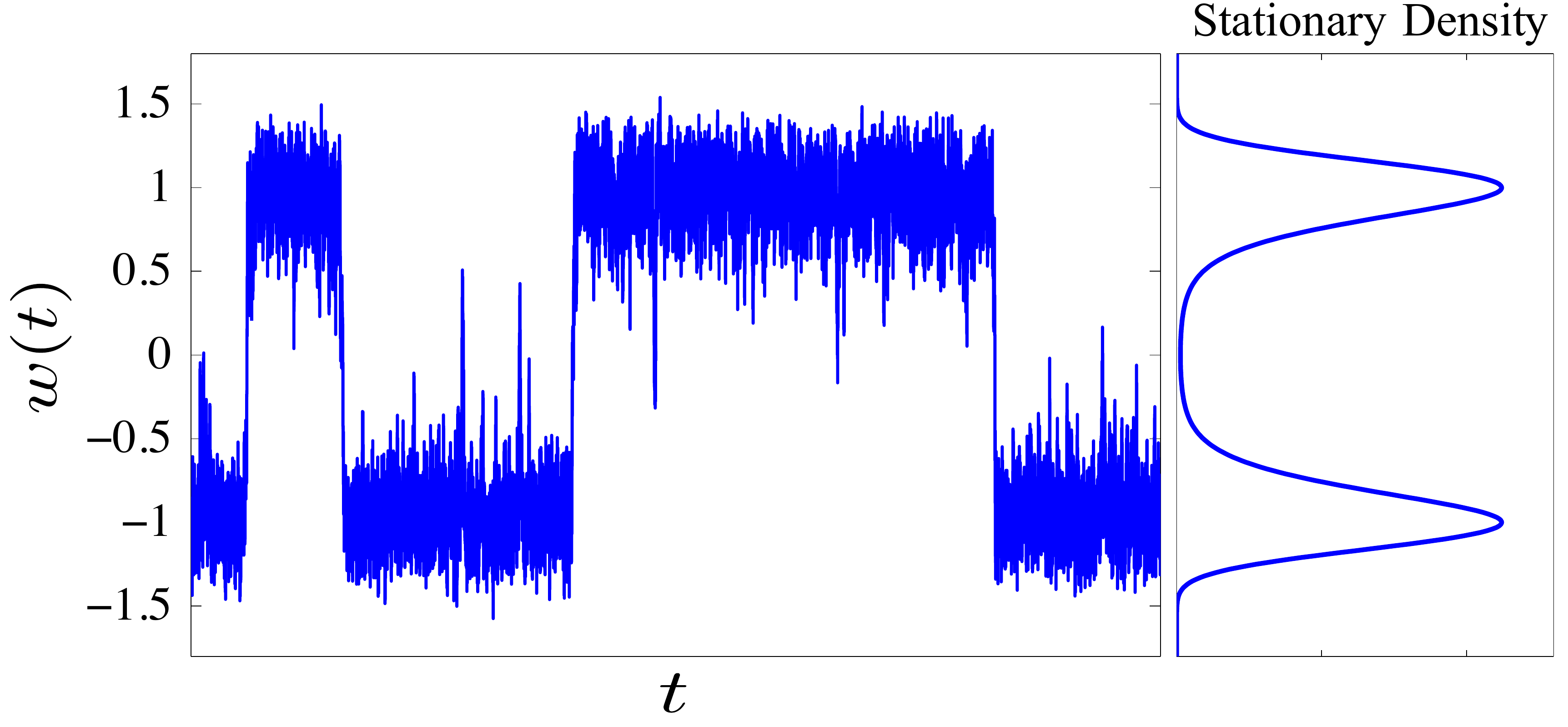}}
\end{center}
\caption{{\bf Bistable switching.} The left panel shows a stochastic process $w(t)$ that spends most of its time near the deterministic stable fixed points $\pm 1$.  The right panel shows the stationary density for this process which has two distinct peaks. }\label{fig:bistab_ex}
\end{figure}

Bistable switching behavior has been observed in variety of physical systems, including gene networks \cite{lai2004,gardner2000}, neural systems \cite{cossart2003,fuster1995}, and neural growth processes \cite{betz2006}. In the simplest example, bistable switching arises when noise is added to a system that has deterministic bistability, i.e., Kramers' problem \cite{gardiner}.  In this case, the stochastic process will randomly switch between the two deterministic fixed points (left panel in Figure \ref{fig:bistab_ex}). The resulting stationary density of the process will have two distinct peaks centered around the deterministic stable fixed points (right panel in Figure \ref{fig:bistab_ex}). In \cite{newby2014}, we demonstrated that similar switching dynamics arise when noise is added to systems with a deterministic limit cycle. Importantly, in the examples studied in \cite{newby2014}, the limit cycle always exists, and thus the points where the stochastic process spent most of its time represent metastable points that do not exist in the deterministic system.  Through the analysis of two simple planar limit cycle systems, we were able to identify precisely when this bistable switching will occur.  However, the underlying dynamics of the bistable behavior in these systems can be quite variable (see Figure \ref{fig:bistabXY}).  Here, we systematically analyze the bistable behavior in these simple systems, and identify the mechanisms leading to different types of switching dynamics.

%


The interaction of the additive noise with the dynamics away from the deterministic limit cycle is the key to understanding the different switching behaviors. The two planar limit cycle systems we analyze have areas of the vector field where the rotation occurs in the opposite direction of the limit cycle.  However, the two systems differ in the placement of this counterrotating behavior.  In one system, the counterrotation occurs on only one side of the limit cycle, whereas in the other system, the limit cycle is surrounded on both sides by a counterrotating vector field.  We will show that these different vector fields lead to very different bistable switching behaviors when perturbed by white noise.  In one case, the noise gets amplified by the vector field, and switching between metastable states occurs frequently.  However, in the other case, the noise gets suppressed, and switching between metastable states occurs very rarely. Through the use of stochastic phase reduction methods \cite{teramae2009, yoshimura2008}, we show that the directionality of switching is controlled by the drift of the phase reduced system while the rate of switching, similar to the Kramers' problem, is determined by the magnitude of the diffusivity relative to the drift.  Importantly, this work illustrates that a wide range of bistable switching dynamics can be modelled by limit cycle systems with additive noise, and provides further evidence that noise can affect seemingly similar nonlinear systems in strikingly different ways.

The paper is organized as follows.  First we present the equations for the two planar limit cycle systems we analyze.  Next, we give the details of the transformation of the system to asymptotic phase which then allows us to reduce the two dimensional systems to a single stochastic differential equation for the evolution of the phase variable.  We then review how the onset of bistable switching behavior can be predicted from the phase model by examining the drift term in the conservation (Fickian) form of the stationary Fokker-Planck equation.  We then illustrate that even though the two systems display qualitatively similar stationary densities, the dynamics of the bistable switching in each system is quite different. By deriving asymptotic approximations to the stationary flux, we show that the directionality of the switching is determined by the sign of the flux.  Lastly, by exploring the terms in the reduced phase system, we show that the rate of switching is determined by the relative magnitude of the drift and diffusivity of the stochastic process.


\section{Models}
\label{sec:models}
To explore this bistable switching behavior in limit cycle sytems, we consider the following analytically tractable deterministic oscillator 

\begin{eqnarray}
  \label{eq:XY}
  \dot{x} &=& -\omega y + \gamma x(1-\rho^{2}) + c \gamma y Q(\rho) \equiv F(x,y) \\
  \dot{y} &=& \omega x + \gamma y(1-\rho^{2}) - c\gamma x Q(\rho)\equiv G(x,y), \nonumber
\end{eqnarray}
where $\rho=\sqrt{x^2+y^2}$ and the function $Q(\rho)$ is such that $Q(1) = 0$ and determines the rotation away from the limit cycle.  The dynamics of the system are easier to understand if \eqref{eq:XY} is rewritten in polar coordinates 

\begin{eqnarray}
  \label{eq:polar}
  &&\dot{\theta} = \omega - \gamma c Q(\rho)\\
   &&\dot{\rho} = -\gamma \rho(\rho^{2} - 1) \nonumber.
\end{eqnarray} 
In particular, it can now easily be seen how $Q(\rho)$ affects the rotation away from the limit cycle at $\rho=1$.  We assume that the limit cycle is strongly attracting so that $\gamma \gg \omega$ is a large parameter. If $c = 0$, then \eqref{eq:polar} is independent of $Q$ and yields the radial isochron clock model when $\gamma=1$.  In this case, the rotation $\dot{\theta}$ is constant away from the limit cycle and independent of $\rho$.
If $c \neq 0$, then the rotation changes direction for values of $\rho=\rho_{*}$ such that $Q(\rho_{*}) = \omega/(c\gamma)$.
We consider the following two cases:
\begin{eqnarray}
  &&Q_{1}(\rho) = \rho^{2} - 1\label{eq:Q}\\
  &&Q_{2}(\rho) = \omega(1 - \rho)^{2} \nonumber.
\end{eqnarray}
$Q_1$ yields the Stuart-Landau model \cite{stuart1960}.  However, for reasons made clear below, we will refer to the system with $Q_{1}$ as the counterrotating (CR) whereas we will refer to $Q_2$ as the antirotating (AR) model.  It follows that the limit cycle rotates counter clockwise with period $2\pi/\omega$ on $\rho = 1$.

\begin{figure}[h!]
\begin{center}
\hspace{0cm} CR System \hspace{4cm} AR System\\
\subfigure[\hspace{6cm}]{\includegraphics[scale=.3]{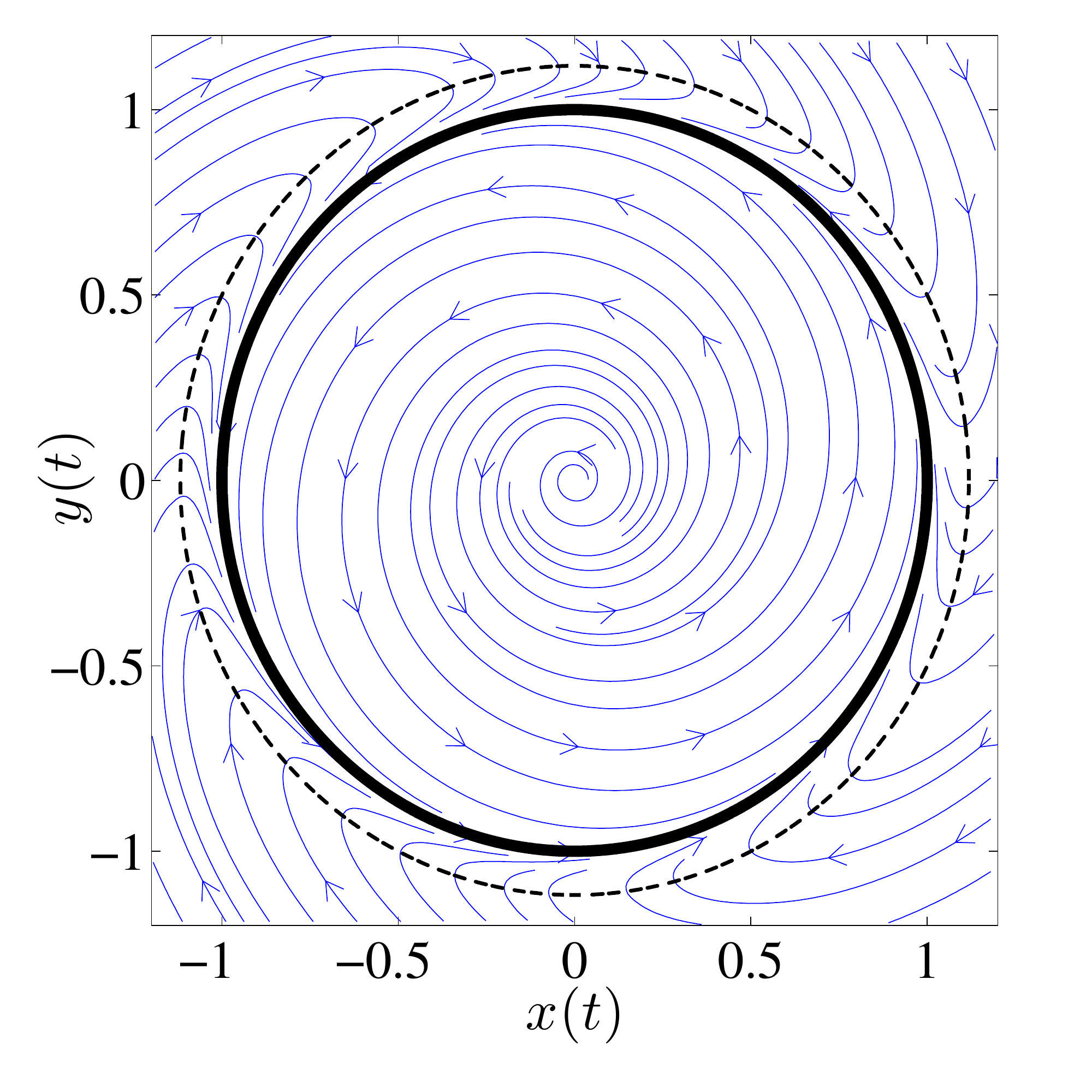}}\subfigure[\hspace{6cm}]{\includegraphics[scale=.3]{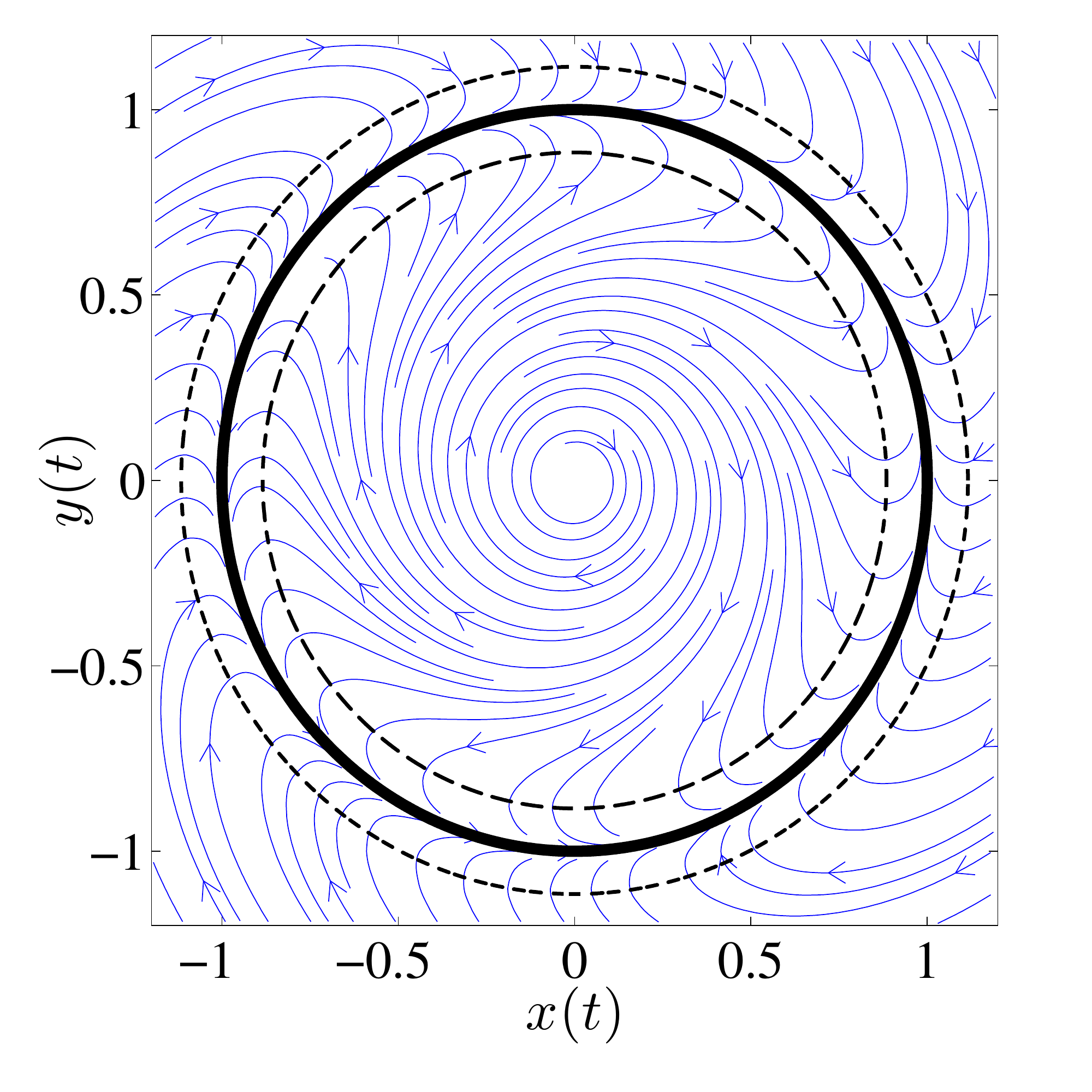}}
\end{center}
\caption{{\bf Vector fields of the two planar systems.} Thin lines represent streamlines of the deterministic vector fields.  Thick line is the limit cycle.  (a) The counterrotating (CR) system with $c=4$ and $\gamma=1$. The vector field changes rotation direction at one value of $\rho$ outside of the limit cycle (dashed line).  (b) The antirotating (AR) system with $c=15$ and $\gamma=5$.  The vector field changes rotation direction at two values of $\rho$ one outside and one inside of the limit cycle (dashed lines). }\label{fig:vectfield}
\end{figure}

Figure \ref{fig:vectfield} plots the vector field of \eqref{eq:XY} for the two different choices of the function $Q(\rho)$ \eqref{eq:Q}. Note that, regardless of the definition of the function $Q(\rho)$, the limit cycle is given by $x(t)=\cos(\omega t)$ and $y(t)=\sin(\omega t)$ (thick black line). For the CR system, there is a unique value of $\rho_{*} = \sqrt{1+\omega/(c\gamma)}$ where the rotation changes sign (dashed line); if $c > 0$ then $\rho_{*}>1$ and if $c < 0$ then $\rho_{*} <1$ (not shown). On the other hand, for the AR system, we have that $\rho_{*} = 1 \pm \sqrt{1/(c\gamma)}$. Hence, there are two values of $\rho_{*}$ if $c>0$ (one inside the limit cycle and one outside, dashed lines) and none if $c<0$.  For simplicity, we only consider the $c>0$ case for both systems. Note that for both the CR and AR systems, increases in $c$ and $\gamma$ cause $\rho^*$ to move closer to the limit cycle.

Next, we add independent white noise to each component
\begin{eqnarray}
  \label{eq:XYnoise}
  \dot{x} &=& F(x,y) + \sigma w_x \xi_{x}(t) \\
  \dot{y} &=& G(x,y) + \sigma w_y \xi_{y}(t), \nonumber
\end{eqnarray}
where $\sigma$ is the noise strength, $w_{x}$ and $w_{y}$ are constants, and $\xi_i(t)$, $i=x,y$, is  Gaussian white noise with zero mean and unit variance (i.e., $\langle \xi_i(t)\rangle = 0$ and $\langle \xi_i(t)\xi_i(s)\rangle = \delta(t-s)$). If $w_{x} = w_{y}$, then the stochastic process is rotationally symmetric. In this case, the CR system acts to amplify noise fluctuations whereas in the AR system, depending on the value of $c$, the noise can cause the system to rotate in the opposite direction of the deterministic limit cycle or even cease oscillating altogether \cite{newby2014}.  When the statistical rotational symmetry is broken, $w_x \neq w_y$, it was also shown that the CR and AR systems can display bistable switching between two metastable states (\cite{newby2014} and see below).  Since we are interested in further examining this bistable behavior, for the rest of the paper we assume, without loss of generality, that $w_x=0$ and $w_y=1$.  This simplifies \eqref{eq:XYnoise} to the following system of stochastic differential equations (SDEs)
\begin{eqnarray}
  \label{eq:XYbistab}
  \dot{x} &=& F(x,y) \\
  \dot{y} &=& G(x,y) + \sigma \xi(t), \nonumber
\end{eqnarray}
To simulate the SDEs, we employ the Euler-Mayurama method with a sufficiently small time step.

\section{Bistable Switching in Limit Cycle Systems}

\begin{figure}[h!]
\begin{center}
\hspace{0cm} CR System \hspace{5.5cm} AR System\\
\subfigure[\hspace{6cm}]{\includegraphics[scale=.3]{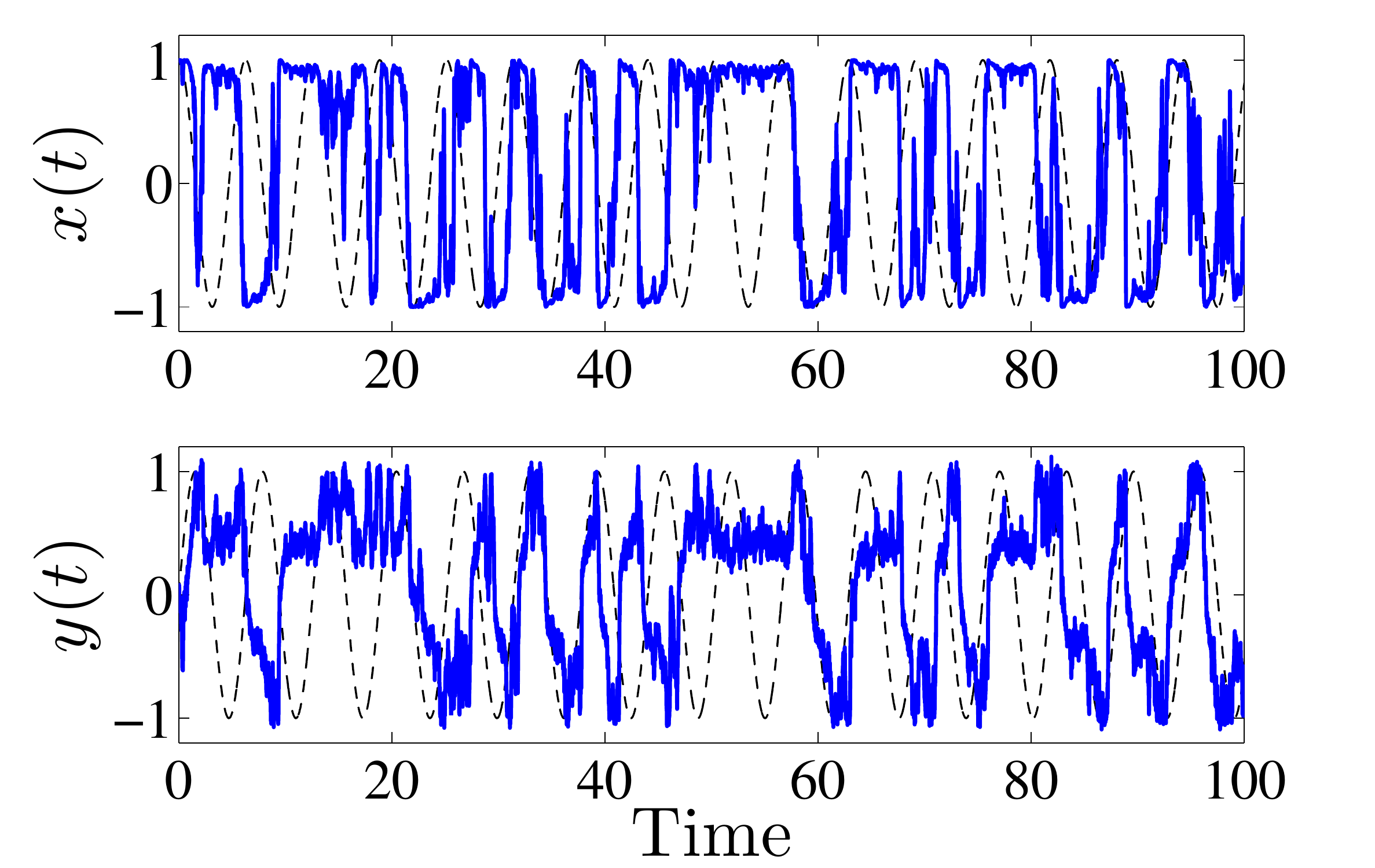}}\subfigure[\hspace{6cm}]{\includegraphics[scale=.3]{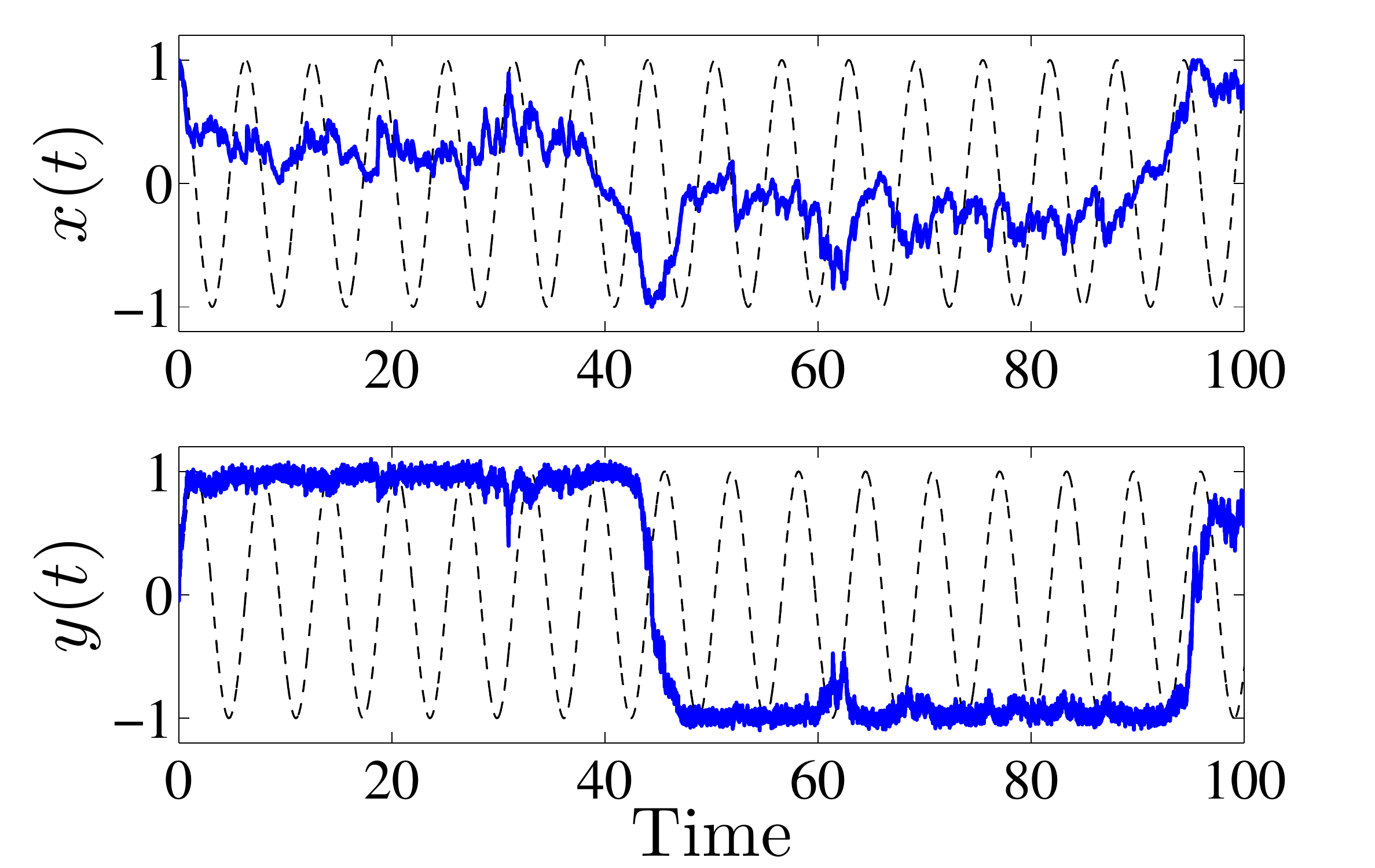}}
\end{center}
\caption{{\bf Bistable switching in the CR and AR systems.} Dashed lines show the deterministic limit cycle $\sigma=0$ (which is the same in both cases). For each system $\gamma = 50$ and $\omega=1$ (a) CR system with $c=4$ and $\sigma=0.5$. Bistable switching is most prominent in the x-component.  (b) AR system with $c=20$ and $\sigma=0.45$. Bistable switching is most prominent in the y-component. }\label{fig:bistabXY}
\end{figure}

In previous work, it was shown that with $c \approx 1$ and moderate levels of noise $\sigma$, the two planar limit cycle systems exhibit bistable-like switching.  Figure \ref{fig:bistabXY} plots example trajectories for both systems. For the CR system, the $x$ coordinate appears to display strong bistable switching between $\pm 1$ at a fairly frequent rate. However, for the AR system, it is the $y$ coordinate that appears most like a bistable switch, but does so at a rate much slower than that of the CR system.  This shows that the vector field away from the limit cycle interacts with the noise in interesting ways in order to produce this behavior.  Thus, we seek to understand (i) how this bistable behavior emerges in each of these systems, and (ii) why bistable switching is different between the two systems.  To do so, we first utilize a phase reduction technique for stochastic limit cycle systems \cite{teramae2009,yoshimura2008} in order to reduce the system to a one-dimensional Ito SDE which describes the position of the system on its underlying determinsitic limit cycle.

\subsection{Reduction to a Phase Model}

In the absence of noise both planar sytems \eqref{eq:XY} exhibit stable limit cycle oscillations. Since we also assume that the rate of attraction back to the limit cycle is large, i.e., $\gamma \gg \omega$, this places us in the ideal situation use so-called phase reduction techniques \cite{ermentroutbook,kuramoto,teramae2009,yoshimura2008}. In this case, the dynamics of a single phase variable $\varphi$ that denotes the location of a trajectory on the underlying limit cycle will be an accurate (to order $1/\gamma$) description of the full dynamics of the system.  

Consider the noiseless system \eqref{eq:XY} which has a stable $2\pi/\omega$-periodic solution. We define a phase $\varphi\in [0,2\pi/\omega)$ coordinate on the limit cycle solution such that it increases by $2\pi/\omega$ for every cycle of the system. That is, $\varphi$ is defined by 

\begin{equation}
\frac{d\varphi} {dt}=\frac{\partial\varphi} {\partial x} \frac{dx} {dt} + \frac{\partial\varphi} {\partial y} \frac{dy} {dt} \equiv 1\label{eq:phasedef}.
\end{equation}
We use phase instead of polar coordinates because, in the absence of external input, the phase variable $\varphi$ increases monotonically with a constant rate. Thus, any change to the rate of increase of the phase variable will solely be due to the external input (in this case noise) and not due to the intrinsic dynamics of the system. This is not the case for the angle variable $\theta$ in polar coordinates \eqref{eq:polar}.

In general, it is not possible to find the exact mapping from Cartesian coordinates to $\varphi$.  However, for the planar systems we use here, the mapping $(x,y)\mapsto \varphi$ can be found analytically.  To see this, recall the polar coordinate form of the systems \eqref{eq:polar}.  Owing to the simplicity of this system, we can easily transform the angle $\theta$ to $\varphi$ using 

\begin{equation*}
\dot{\varphi}=\frac{1} {\omega}\left[\dot{\theta}+\gamma c Q(\rho)\right].
\end{equation*}
Integrating both sides of the above equation with respect to time yields

\begin{eqnarray*}
\varphi &=& \frac{1} {\omega}\left[\theta + \gamma c \int Q(\rho) dt\right]\\
&=& \frac{1} {\omega}\left[\theta + \gamma c \int \frac{Q(\rho)} {-\gamma \rho(\rho^2-1)} d\rho\right].
\end{eqnarray*}
Note that we have ignored the integration constant which is an arbitrary phase shift related to the initial starting point.  Next, we define a new amplitude variable $r=\rho-1$ which is the distance of the system from the deterministic limit cycle. Using the fact that $\theta=\tan^{-1}(y/x)$ and $\rho=\sqrt{x^2+y^2}$, we arrive at our desired mapping

\begin{equation}
\label{eq:phasemap}
\varphi=\frac{1} {\omega}\left[\tan^{-1}(y/x) +c H(r)\right], \,\,\,\, r=\sqrt{x^2+y^2}-1,
\end{equation}   
where $H(r)\equiv \int \frac{Q(r+1)} {-(r+1)((r+1)^2-1)}dr$. It is important to note that the function $H(\cdot)$ depends on the function $Q(\cdot)$.  Thus, the mapping to phase will be different for the CR and AR systems.

Lines of constant $\varphi$ are called isochrons \cite{winfree} -- all deterministic trajectories starting on an isochron converge as $t\to\infty$.  Hence, isochrons encode information about how the phase of a trajectory responds to a perturbation away from the limit cycle.  Indeed, the gradient of this mapping $\frac{\partial \varphi} {\partial i}$, $i=x,y$ is typically referred to as the phase response curve of the $i$-th component of the limit cycle \cite{ermentroutbook,kuramoto,schwemmerandlewisbooktwoc} as it quantifies the change in phase due to a $\delta$-function perturbation at a particular phase on the limit cycle.

With our mapping in hand \eqref{eq:phasemap}, we transform the stochastic Cartesian system \eqref{eq:XYbistab} to $(\varphi,r)$ coordinates 
\begin{eqnarray}
  \label{eq:rphi}
  \dot{\varphi}&=& 1+\dif n(\varphi, r) + \sigma h(\varphi, r) {\xi}(t)\\
  \dot{r}&=&\gamma f(r) + \dif n_r(\varphi, r) + \sigma g(\varphi, r) {\xi}(t),\nonumber
\end{eqnarray}
\noindent where,

\begin{eqnarray*}
&&h(\varphi,r)\equiv \frac{\partial \varphi} {\partial y} = \frac{\sqrt{1+\lambda^2}} {\omega(r+1)}\sin(\alpha+\psi), \,\,\,\,\,\,\, g(\varphi,r)\equiv \frac{\partial r} {\partial y} = \sin(\alpha)\\
&& f(r)\equiv \frac{\partial r} {\partial x} F(x(\varphi,r),y(\varphi,r))+\frac{\partial r} {\partial y} G(x(\varphi,r),y(\varphi,r)) = -r(r+1)(r+2)\\
&& \lambda \equiv c(r+1)H'(r), \,\,\,\,\,\,\, \psi \equiv \tan^{-1}(\lambda), \,\,\,\,\,\,\, \alpha \equiv \omega \varphi + cH(r),
\end{eqnarray*}
and we have used the inverse mappings $x(\varphi,r)=(r+1)\cos(\alpha)$ and $y(\varphi,r)=(r+1)\sin(\alpha)$. While the noise is additive in cartesian coordinates \eqref{eq:XYbistab}, it is multiplicative in $(\varphi, r)$ coordinates. As such, we use the Ito interpretation. The terms that arise from the stochastic change of variables (i.e., the terms that come from the Ito calculus \cite{gardiner}) are $n(\varphi, r) \equiv h(\varphi, r) \frac{\partial h} {\partial \varphi} + {g}(\varphi, r)\frac{\partial h} {\partial r}$ and $n_r(\varphi, r) \equiv {h}(\varphi, r) \frac{\partial g} {\partial \varphi} + {g}(\varphi, r)\frac{\partial g} {\partial r}$.
We stress that \eqref{eq:rphi} is an exact transformation, and no approximations have been used up to this point.  Thus, our goal now remains to project out the variable $r$ in order to arrive at a single differential equation for the phase. 

Assuming that the deterministic limit cycle is highly attracting (i.e., $\gamma \gg \omega$), the system should cling tighly to the underlying deterministic limit cycle, and on average $r$ will be approximately $0$. Thus, we can set $r=0$ in \eqref{eq:rphi} to obtain (to leading order in $\gamma^{-1}$) the one dimensional approximation

\begin{equation}
  \label{eq:phase}
  \dot{\varphi}\sim 1+\dif n_{0}(\varphi) + \sigma h_{0}(\varphi) \xi(t), \quad \gamma \gg \omega,
\end{equation}
where $n_{0}(\varphi) \equiv n(\varphi, 0)$ and $h_{0}(\varphi) \equiv h(\varphi, 0)$ is the phase response curve of the y-component of the deterministic limit cycle.  Note that \eqref{eq:phase} can be rigorously derived by using adiabatic (or quasi-steady state) reduction techniques \cite{gardiner,teramae2009}.  

The stationary density $u_{ss}(\varphi)$ of the reduced system \eqref{eq:phase} will provide insight into the dynamics of the full model.  The stationary density obeys the following Fokker-Planck (FP) equation

\begin{equation}
\label{eq:phaseFPIto}
0 = \partial_{\varphi}\left[\partial_{\varphi}\{D_{\varphi}(\varphi)u_{ss}\}-I_{\varphi}(\varphi)u_{ss}\right],
\end{equation}
with periodic boundary conditions. Note that in general analytical solutions to the above FP equation are not possible and we instead find approximate solutions using a forward or backward finite-difference scheme (depending on parameters). $I_{\varphi}(\varphi)\equiv 1+\dif n_{0}(\varphi)$ is the Ito drift, or the deterministic portion of the SDE \eqref{eq:phase}, and $D_{\varphi}(\varphi)\equiv \frac{\sigma^2} {2} h_0^2(\varphi)$ is the diffusivity. The FP equation \eqref{eq:phaseFPIto} can also be written in conservation or Fickian form

\begin{equation}
\label{eq:phaseFPFick}
0 = \partial_{\varphi}\left[D_{\varphi}(\varphi)\partial_{\varphi}u_{ss}-V_{\varphi}(\varphi)u_{ss}\right],
\end{equation}
where $V_{\varphi}(\varphi) \equiv I_{\varphi}(\varphi) - D_{\varphi}'(\varphi)$ is the Fickian drift.  

The Fickian drift is closely related to the stationary density of the the system.  To see this, integrate both sides of \eqref{eq:phaseFPFick} to obtain

\begin{equation*}
-J = D_{\varphi}(\varphi)\partial_{\varphi}u_{ss}-V_{\varphi}(\varphi)u_{ss}, 
\end{equation*}
where $J$ is the probability flux.  If the flux is zero, then $u_{ss}$ has the solution $u_{ss}(\varphi)=Ne^{-\Phi(\varphi)}$ where 

\begin{equation*}
\Phi(\varphi)=\int_0^{\varphi} \frac{-V(\tilde{\varphi})} {D(\tilde{\varphi})}d\tilde{\varphi},
\end{equation*}
and $N$ is a normalization factor.  

On the other hand, the Ito drift represents the average increment of the stochastic process during an infinitesimal time $dt$.  For a periodic system such as ours, the Ito drift is also related to the stationary probability flux.  To see this, integrate both sides of \eqref{eq:phaseFPIto} to obtain

\begin{equation*}
-J = \partial_{\varphi}\left(D_{\varphi}(\varphi)u_{ss}\right)-I_{\varphi}(\varphi)u_{ss}, 
\end{equation*}
where $J$ is again the probability flux. Lastly, integrating both sides of the above equation from $0$ to $2\pi/\omega$ (and noting that $D_{\varphi}(\varphi)$ and $u_{ss}$ are $2\pi/\omega$ periodic) yields

\begin{equation}
J=\frac{1} {2\pi/\omega}\langle I_{\varphi}(\varphi)\rangle_{\varphi} \equiv \frac{1} {2\pi/\omega}\int_0^{2\pi/\omega}I_{\varphi}(\varphi)u_{ss}(\varphi)d\varphi. \label{eq:flux}
\end{equation}
In the case that the stationary density is uniform $u_{ss}=(2\pi/\omega)^{-1}$, the probability flux is proportional to the average Ito drift over one period of the oscillations.  However, in general the stationary density will not be uniform, and its shape will have a large influence on the flux. 


The flux can also be thought of as the average frequency of the oscillations. In the noiseless system, $I_{\varphi}(\varphi)=1$, and thus the flux is $(2\pi/\omega)^{-1}$, which is the frequency of the deterministic limit cycle. In fact, for stochastically perturbed periodic systems that have an Ito drift of $1$, the stationary flux is always $(2\pi/\omega)^{-1}$ regardless of the noise level.


\subsection{Diagnosing Bistability in the CR and AR Systems}

\begin{figure}[h!]
\begin{center}
\hspace{0cm} CR System \hspace{5.5cm} AR System\\
\subfigure[\hspace{6cm}]{\includegraphics[scale=.3]{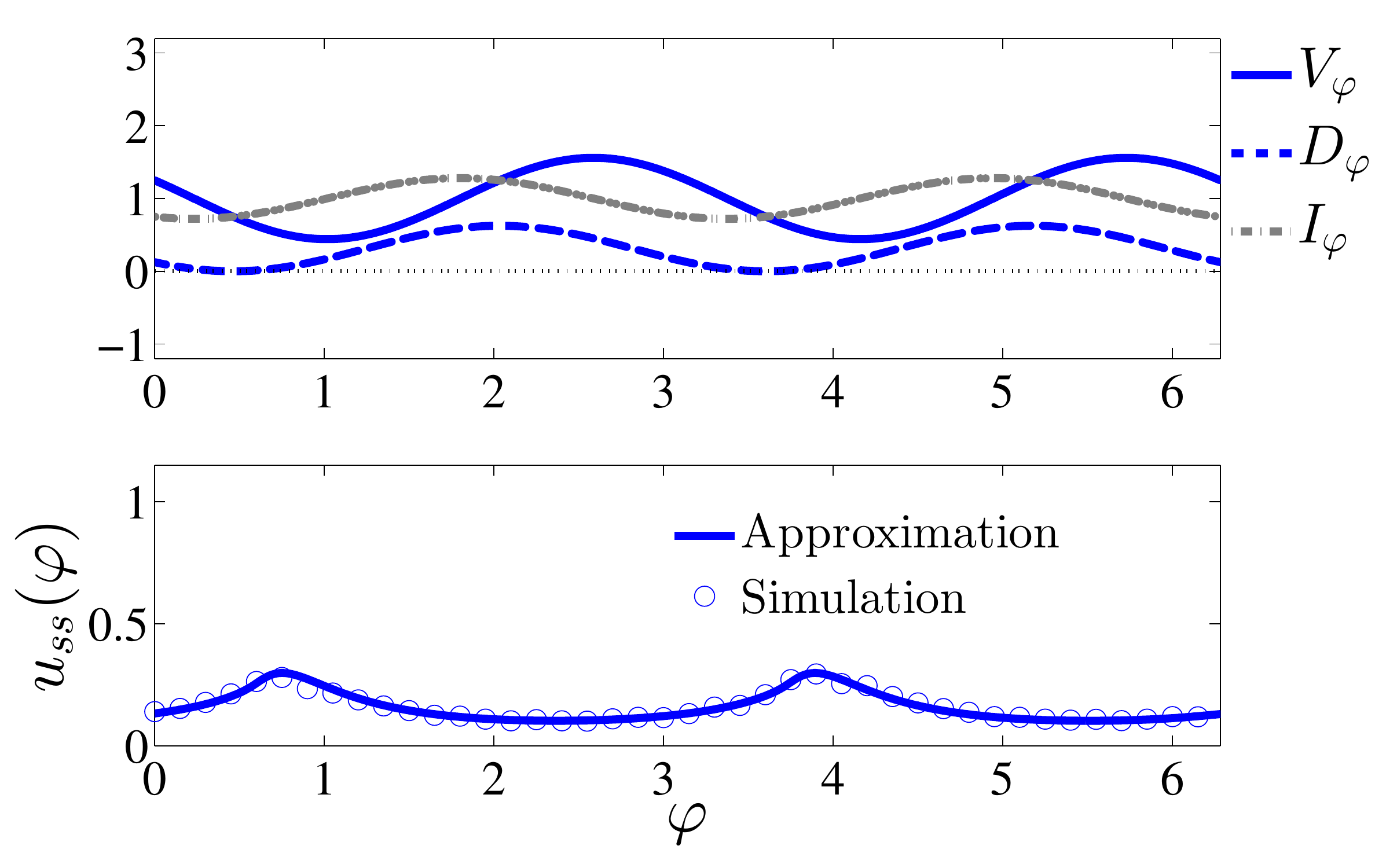}}\subfigure[\hspace{6cm}]{\includegraphics[scale=.3]{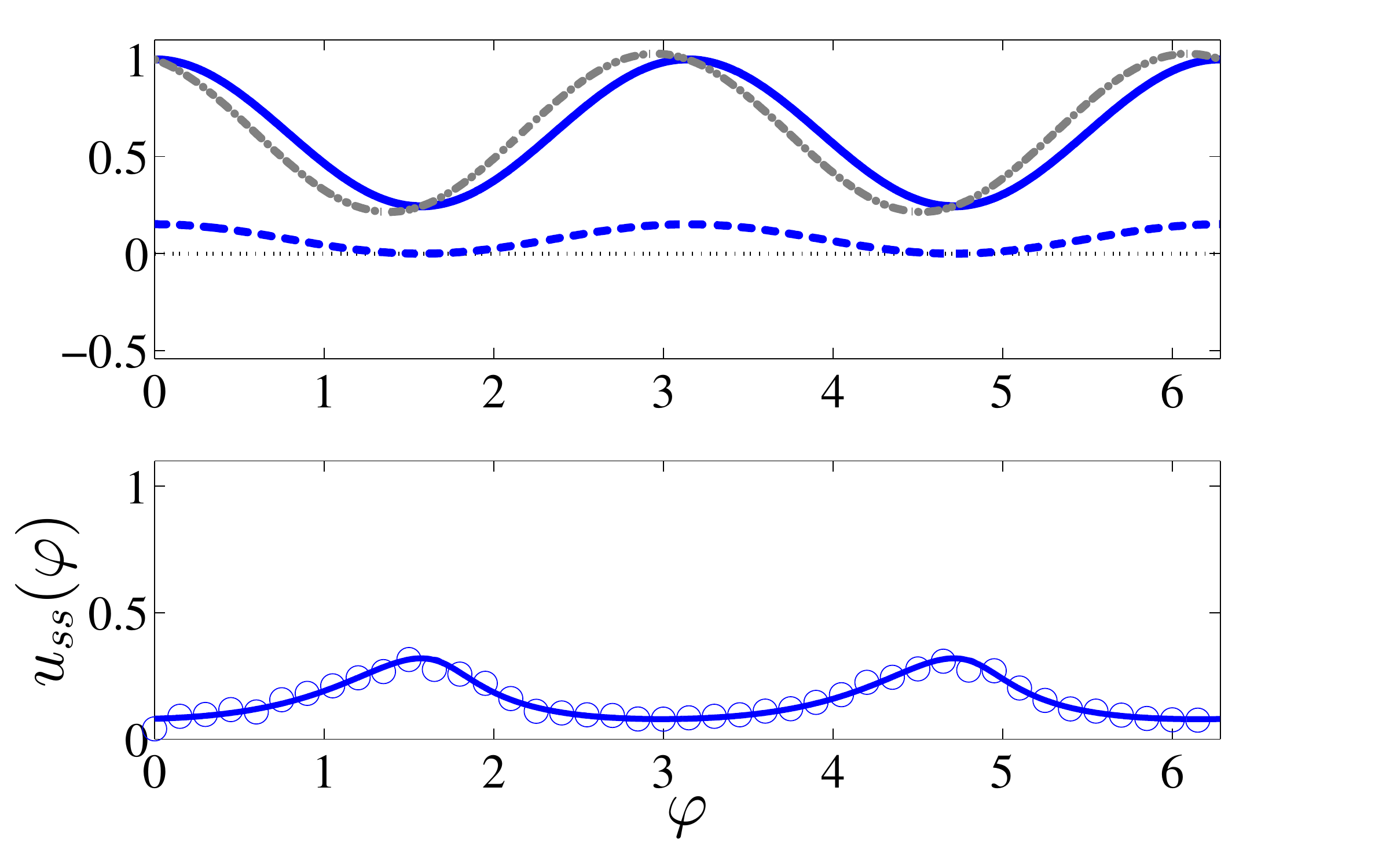}}\\
\subfigure[\hspace{6cm}]{\includegraphics[scale=.3]{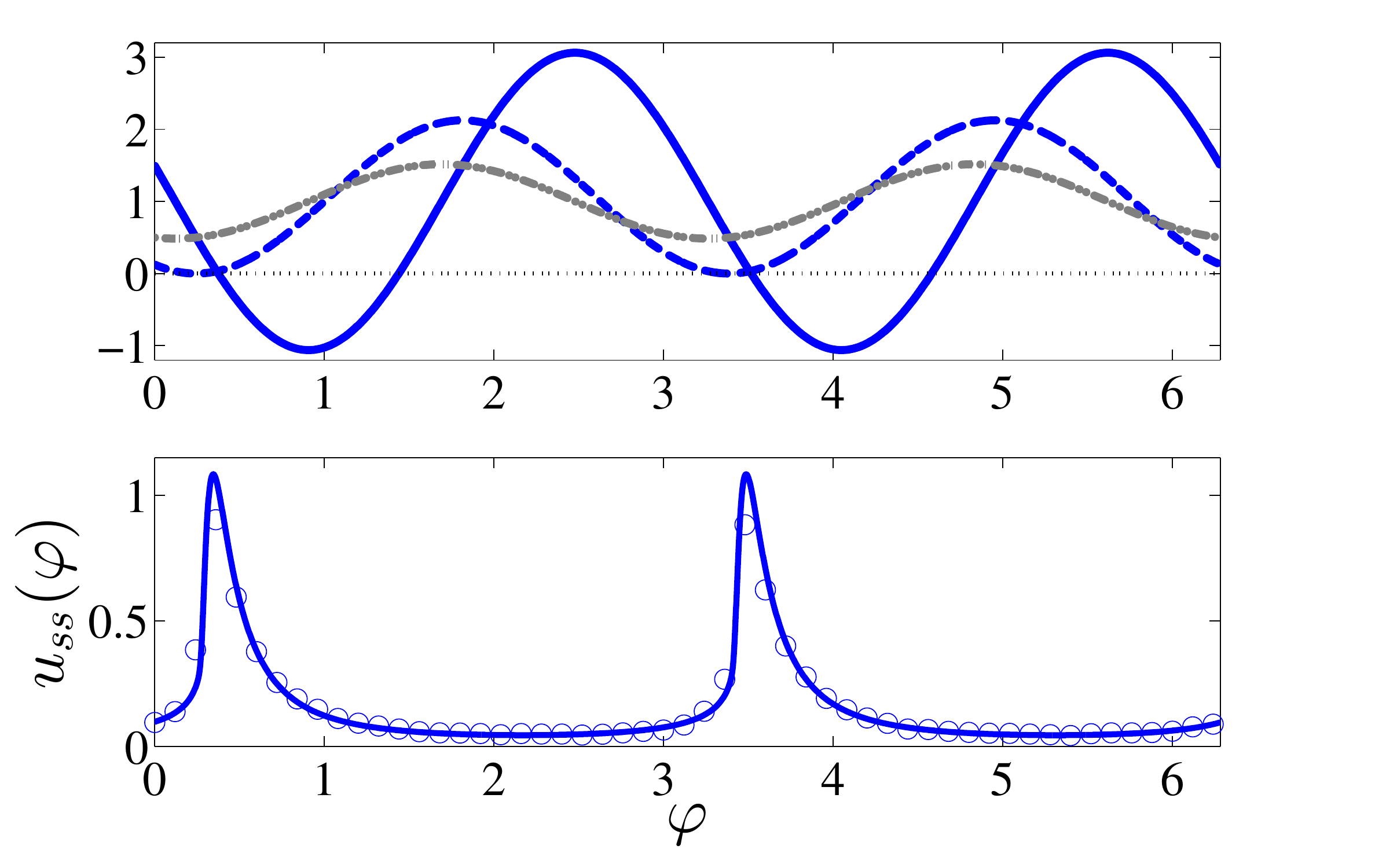}}\subfigure[\hspace{6cm}]{\includegraphics[scale=.3]{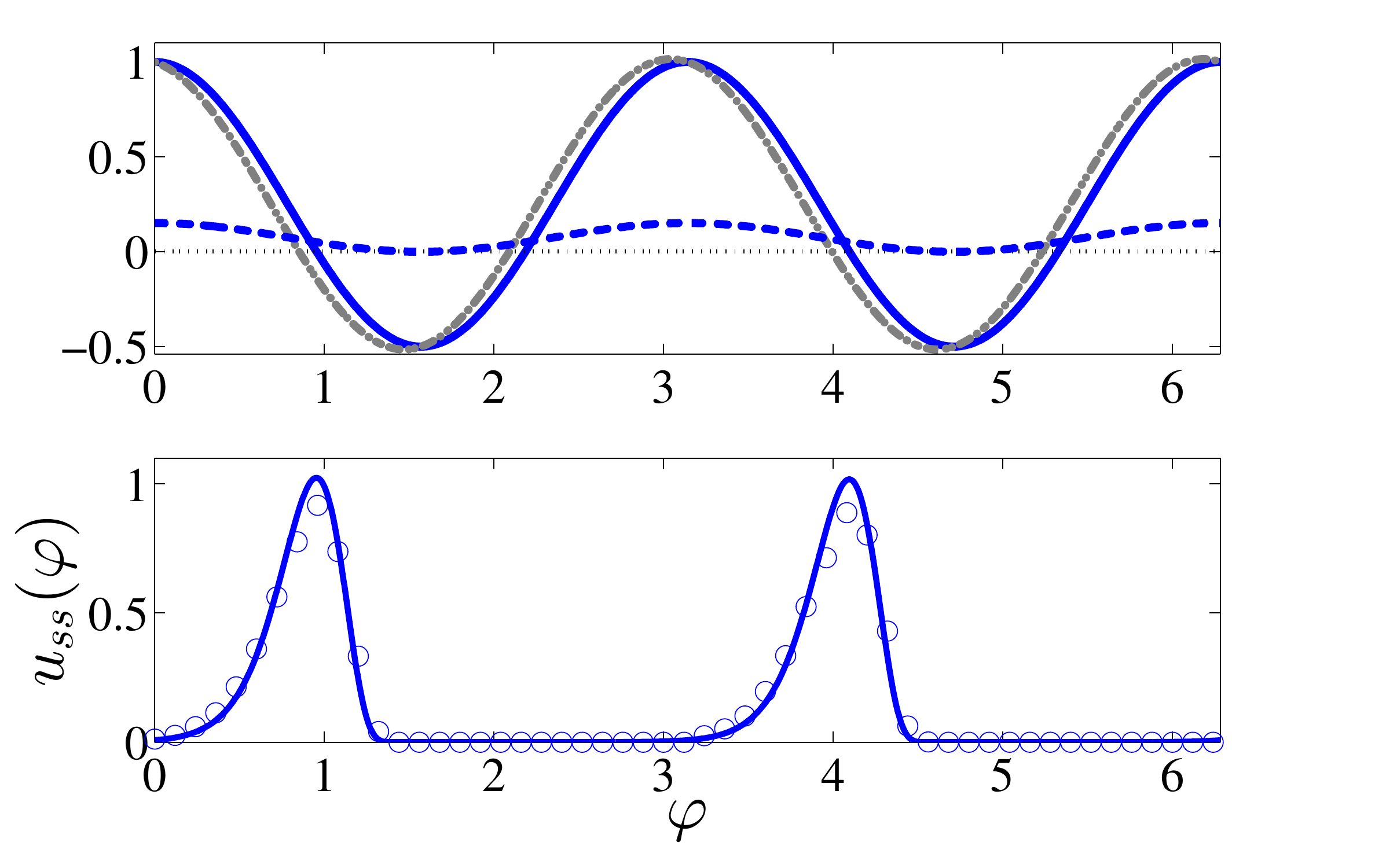}}
\end{center}
\caption{{\bf Bistability occurs when the Fickian drift becomes negative.}  Upper panels plot the  Fickian drift ($V_{\varphi}$, solid lines), Ito drift ($I_{\varphi}$, grey dash-dotted lines), and diffusivity ($D_{\varphi}$, solid lines) for the phase reduced system \eqref{eq:phase}.  Lower panels plot the stationay density of the phase reduced system (solid lines) obtained from solving \eqref{eq:phaseFPIto} along with Monte-Carlo simulations of the full system \eqref{eq:rphi} (open circles) with $\gamma=50$. In all cases, $\omega=1$. The (a) CR system with $c=2$ and $\sigma=0.5$ and the (b) AR system with $c=10$ and $\sigma=0.55$ both have relatively flat stationary densities. Strong peaks in the stationary densities occur when the Fickian drift crosses zero with a negative slope in (c) CR system with $c=4$ and $\sigma=0.5$ and (d) CR system with $c=20$ and $\sigma=0.5$ }\label{fig:driftsdensities}
\end{figure}

Before showing how the onset of bistable switching behavior can be predicted using the phase reduced system, we first give the details of the drift and diffusivity functions which will be vital to our later analysis.  For the CR system, the Ito drift, Fickian drift, and diffusivity are given by   

\begin{eqnarray}
\label{eq:CRstuff}
&&I^{CR}_{\varphi}(\varphi) = 1-\frac{\sigma^2} {2\omega}[c\cos(2\omega\varphi)+\sin(2\omega\varphi)] \\
&&V^{CR}_{\varphi}(\varphi) = 1+\frac{\sigma^2} {2\omega}[-c^2\sin(2\omega\varphi)+c\cos(2\omega\varphi)]\nonumber \\
&&D^{CR}_{\varphi}(\varphi) = \frac{\sigma^2} {2\omega} [c^2\sin^2(\omega \varphi)-c\sin(2\omega\varphi)+\cos^2(\omega\varphi)]\nonumber.
\end{eqnarray}
The corresponding quantities for the AR system are 

\begin{eqnarray}
\label{eq:ARstuff}
&&I^{AR}_{\varphi}(\varphi) = 1-\frac{\sigma^2} {2}\left[\frac{c} {2}\sin^2(\omega\varphi)+\frac{1} {\omega}\sin(2\omega\varphi)\right] \\
&&V^{AR}_{\varphi}(\varphi) = 1-\frac{\sigma^2} {4} c \sin^2(\omega\varphi)\nonumber\\
&&D^{AR}_{\varphi}(\varphi) = \frac{\sigma^2} {2\omega} \cos^2(\omega\varphi)\nonumber.
\end{eqnarray}
Figure \ref{fig:driftsdensities} plots the above quantities along with the stationary density of the phase reduced system (numerical solution of equation \eqref{eq:phaseFPFick}) and the stationary density computed from Monte-Carlo simulations of the full system \eqref{eq:rphi}.  Note the excellent agreement between the phase reduction and the full system (compare open circle to the solid lines in the lower panels). The CR (AR) system is shown in Fig. \ref{fig:driftsdensities} (a) and (c) ((b) and (d)) with two different values for $c$ and all other parameters held constant.  Recall from section \ref{sec:models} that increases in the parameter $c$ causes the oppositely-rotating region(s) of the vector field to move closer to the limit cycle (i.e., increasing $c$ brings $\rho_*$ closer to $1$). It can clearly be seen that, in both systems, increasing the value of $c$ also causes the Fickian drift to become negative which signals the onset of the bistable switching behavior.  Furthermore, the two points where $V_{\varphi}$ changes sign from positive to negative (see upper panels of  Fig.~\ref{fig:driftsdensities} (c) and (d)) as $\varphi$ is increased are local maxima of the stationary density, where the oscillator spends most of its time, struggling to move past a dynamical barrier imposed by the noise. Thus, the zeros of the Fickian drift that have negative slope are \emph{metastable} points that the system switches between. It is important to note that the noise-induced contributions  to the Fickian drift (the $\sigma^2$ terms) are what causes it to become negative.  Indeed, if $\sigma$ is close to zero, then $V_{\varphi}\approx 1$ for both systems.


We can already see interesting differences between the two systems emerge in their respective phase reductions.  In the AR system, the diffusivity is independent of $c$ which is not the case for the CR system. Thus, in the CR system, changing the dynamics of the vector field around the limit cycle (by varying $c$) affects both the drift and diffusivity of the phase reduced system, whereas in the AR system, the fluctuations of the phase variable are unaffected by the dynamics off the limit cycle (compare the dashed traces).  It can also be seen that in the AR system, the diffusivity is small compared to the magnitude of the drift (compare the dashed lines to the solid lines in the upper panels of Fig.~\ref{fig:driftsdensities} (b) and (d)).  Lastly, in the AR system, the the Ito and Fickian drifts are approximately the same whereas in the SL system, they are quite different (compare the dash dotted lines to the solid lines in the upper panels of Fig.~\ref{fig:driftsdensities}).  

\begin{figure}[h!]
\begin{center}
\subfigure[\hspace{6cm}]{\includegraphics[scale=.25]{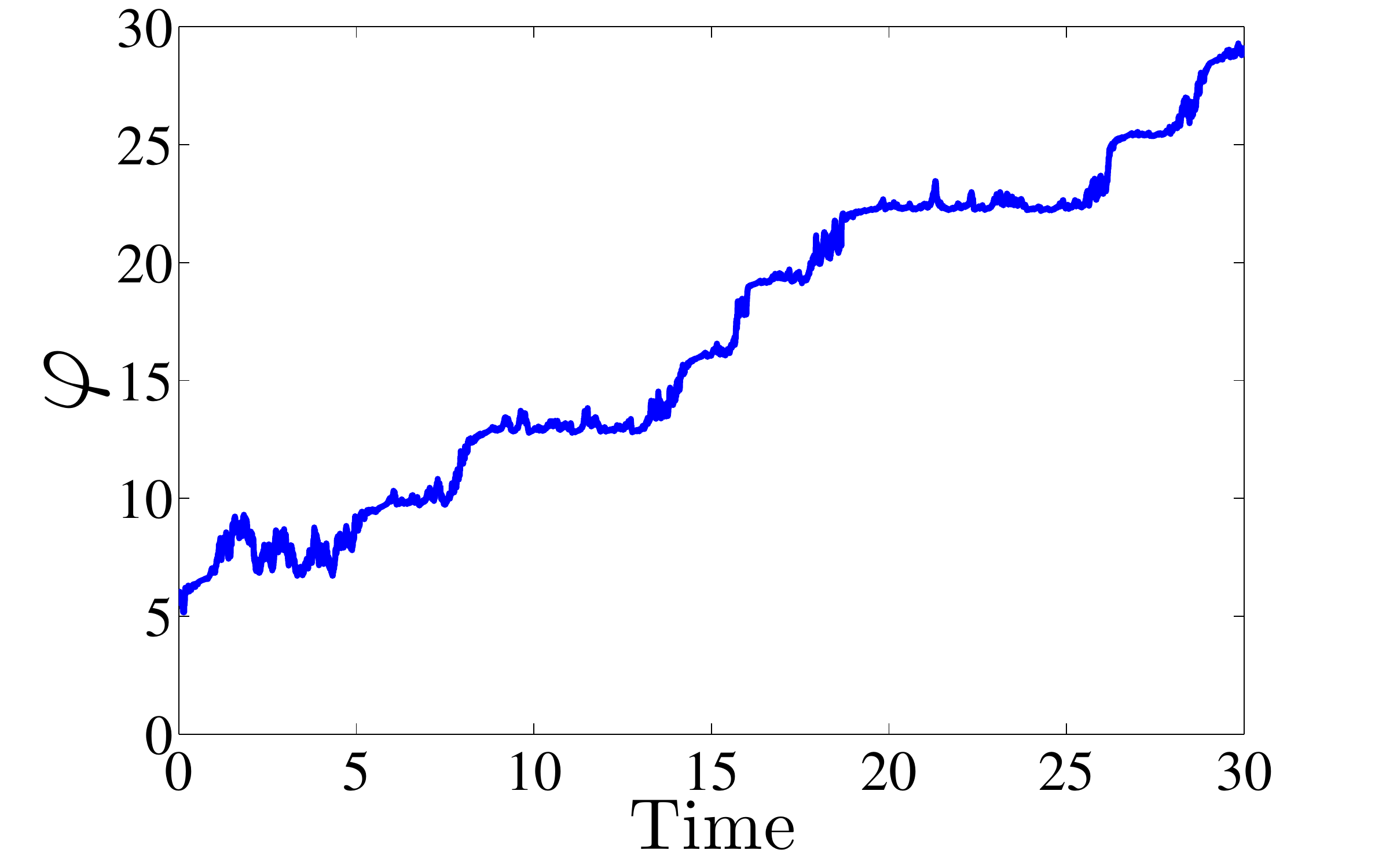}}\subfigure[\hspace{6cm}]{\includegraphics[scale=.25]{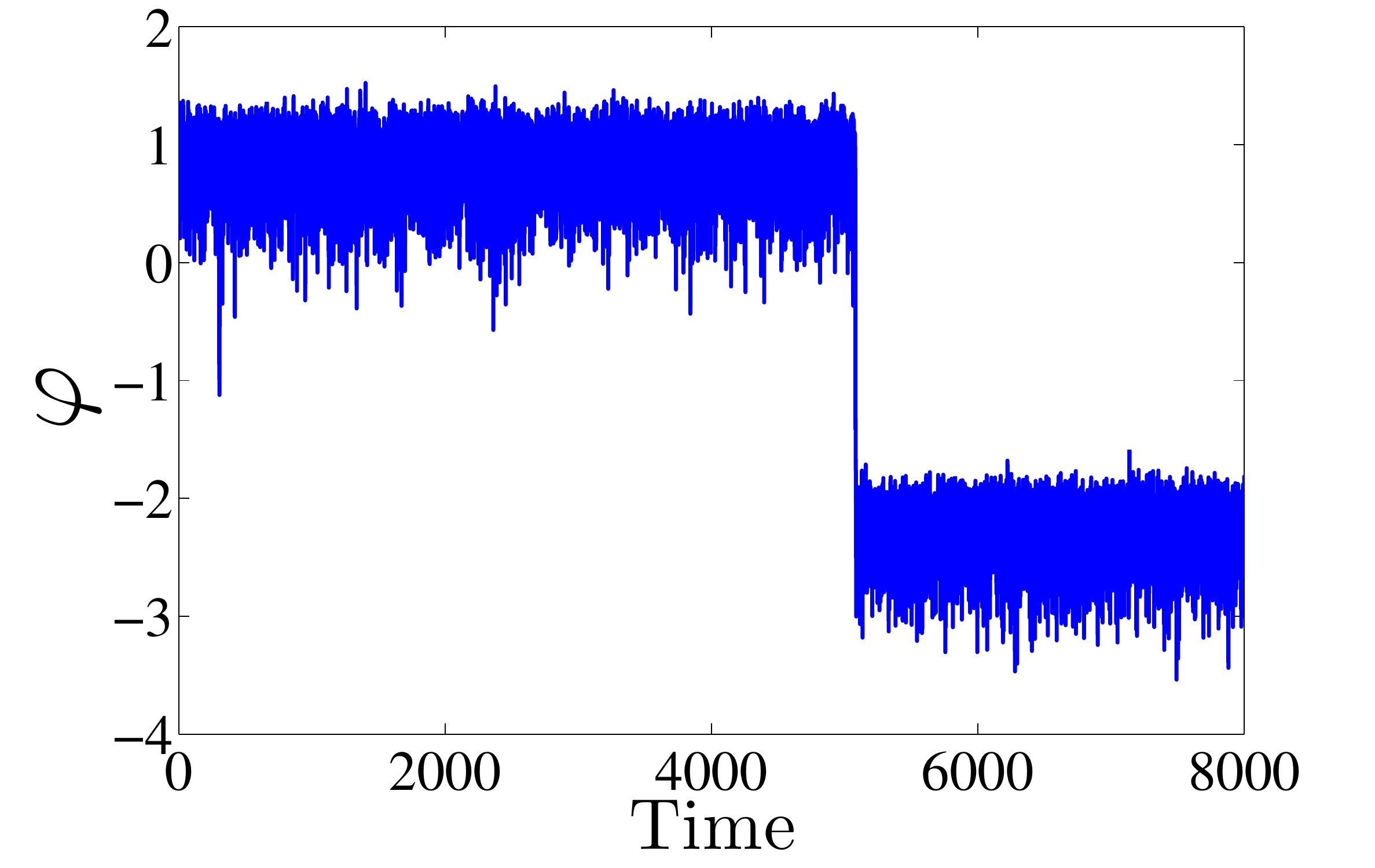}}
\end{center}
\caption{{\bf CR and AR systems display different switching dynamics.}  Sample phase trajectories from simulations of the full system \eqref{eq:rphi} with $\gamma=50$ and $\omega=1$. (a) CR system with $c=4$ and $\sigma=0.5$ displays frequent switching while (b) AR system with $c=20$ and $\sigma=0.55$ displays very rare switiching (note difference in x-axis scale). }\label{fig:bistablephase}
\end{figure}

Thus, the noise interacts with the nonlinear dynamics of the CR and AR systems in different ways to produce stochastic phase variables with qualitatively similar stationary behavior.  However, recalling the trajectories shown in Figure \ref{fig:bistabXY}, and noting the differences in the shapes of the stationary densities in the lower panels of Figure \ref{fig:driftsdensities} (c) and (d), we can begin to see that there may be stronger differences in the transient behavior of the CR and AR systems.  Indeed, if we plot stochastic phase trajectories from simulations of the full system \eqref{eq:rphi} as in Figure \ref{fig:bistablephase}, we can see strong differences emerging between the two systems. The first thing to note is that the switching behavior observed in Cartesian coordinates translates to sudden jumps in the trajectory of $\varphi$ that we will refer to as ``phase jumps''.  The CR system displays fairly frequent phase jumps whereas in the AR system, these transitions are rare (compare the x-axis scale in Fig. \ref{fig:bistablephase} (a) and (b)).  This is similar to what we observed in simulations of the Cartesian system in Figure \ref{fig:bistabXY}.  Secondly, in the CR system, the phase jumps are in the positive phase direction (the same direction as the deterministic limit cycle) but in the AR system the jumps are in the opposite direction of the deterministic limit cycle.

We now seek to understand why the bistable behavior differs between the two systems. We will first concentrate on the different switching directions.  In particular, we will show that the stationary probabilty flux is related to the directionality of the rotation of the oscillators, and we will derive analytical approximations to the flux in order to gain insight into the mechanism underlying the different switching directions.  


\section{Stationary Probability Flux Determines Switching Direction}

Here, we derive analytical approximations to the flux for the AR and CR systems. For both systems, the bistable behavior increases as $c$ and $\sigma$ are varied such that the Fickian drifts becomes more negative. Because of this ambiguity, we introduce different scalings for the two systems in order to arrive at a single parameter which, when increased, causes bistable switching to occur.  With this scaling, we are also able to take a distiguished limit ($c\to \infty$ and $\sigma \to 0$) of the phase reduced systems in order to obtain analytically tractable approximations.  We note that if $c$ becomes too large, then the phase reduction itself may cease being an accurate description of the full system.  However, as long as we assume $\gamma > c \gg 1$, then the results in the next section hold, and we verify this fact with the use of numerical simulations.


\subsection{CR System}

We first introduce the parameters $\varepsilon=\frac{1} {c}$ and $K_{CR}=\frac{\sigma} {\varepsilon}$. Keeping the parameter $K_{CR}$ constant as we simultaneously let $\sigma \to 0$ and $\varepsilon \to 0$ allows us to expand the system in an asymptotic series in $\varepsilon$.  To see this, substitute $c=\frac{1} {\varepsilon}$ and $\sigma=\varepsilon K_{CR}$ into equation \eqref{eq:CRstuff} to obtain

\begin{eqnarray}
\label{eq:CRexpansions}
I^{CR}_{\varphi}(\varphi) &=& 1-\varepsilon\frac{K_{CR}^2} {2\omega}\cos(2\omega\varphi)-\varepsilon^2\frac{K_{CR}^2} {2\omega}\sin(2\omega\varphi) \\
&\equiv&I^{CR,0}_{\varphi}(\varphi)+\varepsilon I^{CR,1}_{\varphi}(\varphi)+\mathcal{O}(\varepsilon^2)\nonumber\\
V^{CR}_{\varphi}(\varphi) &=& 1-\frac{K_{CR}^2} {2\omega}\sin(2\omega\varphi)+\varepsilon\frac{K_{CR}^2} {2\omega}\cos(2\omega\varphi)\nonumber \\
&\equiv&V^{CR,0}_{\varphi}(\varphi)+\varepsilon V^{CR,1}_{\varphi}(\varphi)\nonumber\\
D^{CR}_{\varphi}(\varphi) &=& \frac{K_{CR}^2} {2\omega} \sin^2(\omega \varphi)-\varepsilon\frac{K_{CR}^2} {2\omega}\sin(2\omega\varphi)+\varepsilon^2\frac{K_{CR}^2} {2\omega}\cos^2(\omega\varphi)\nonumber\\
&\equiv&D^{CR,0}_{\varphi}(\varphi)+\varepsilon D^{CR,1}_{\varphi}(\varphi)+\mathcal{O}(\varepsilon^2)\nonumber.
\end{eqnarray}
Next, we expand the stationary density $u_{ss}=u_{ss}^0+u_{ss}^1+\mathcal{O}(\varepsilon^2)$ to obtain an asymptotic approximation to the flux \eqref{eq:flux}

\begin{equation}
\label{eq:CRfluxapprox}
J=\frac{1} {2\pi/\omega}\left[1+\varepsilon\int_0^{2\pi/\omega}I^{CR,1}_{\varphi}(\varphi)u_{ss}^0(\varphi)d\varphi\right]+\mathcal{O}(\varepsilon^2),
\end{equation}
where we used the fact that $I^{CR,0}_{\varphi}(\varphi)=1$, $\int_0^{2\pi/\omega}u_{ss}^0(\varphi)d\varphi=1$, and $\int_0^{2\pi/\omega}u_{ss}^1(\varphi)d\varphi=0$. Thus, in order to close our approximation, we must find the leading order approximation to the stationary density $u_{ss}^0$. 

\begin{figure}[h!]
\begin{center}
\subfigure[\hspace{6cm}]{\includegraphics[scale=.35]{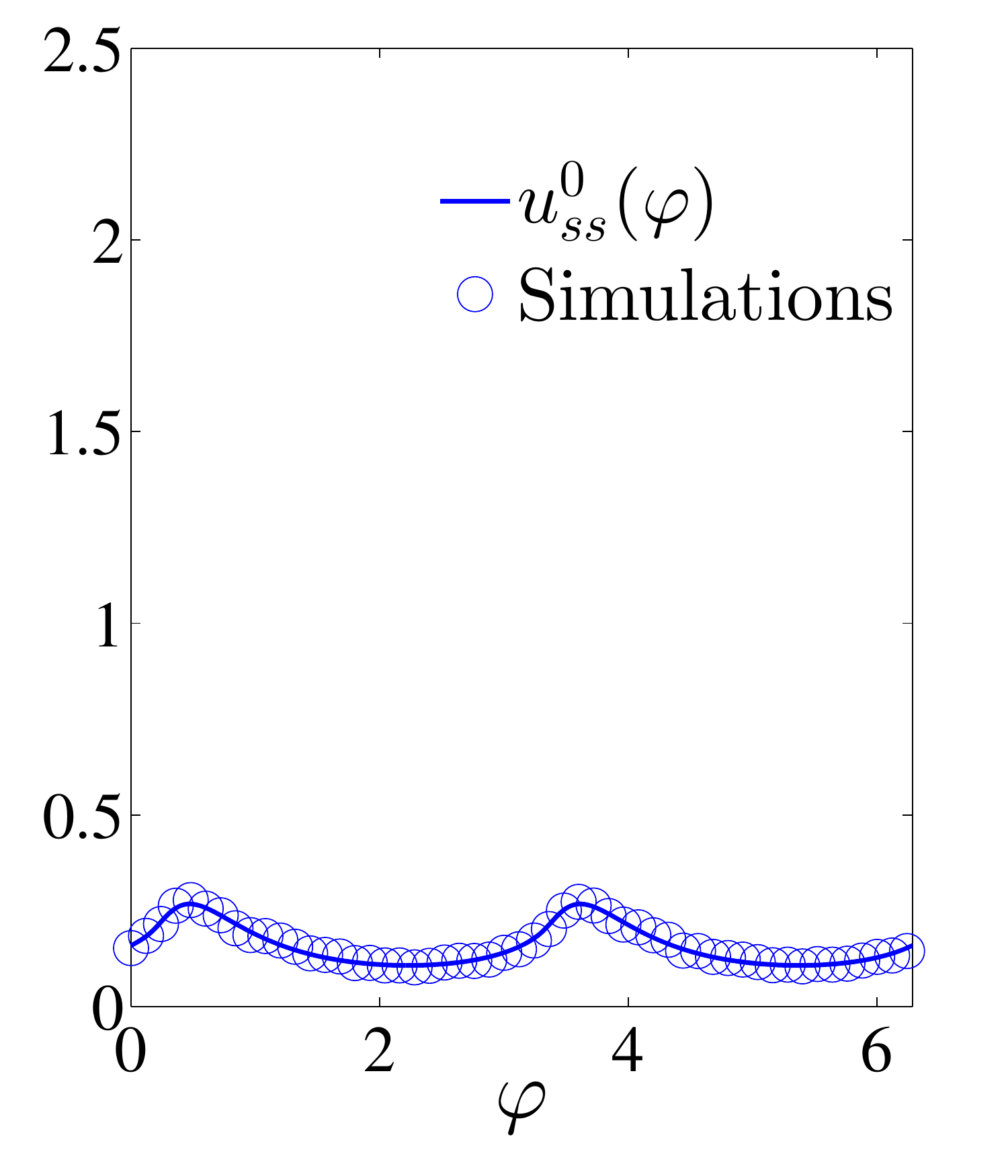}}\subfigure[\hspace{6cm}]{\includegraphics[scale=.35]{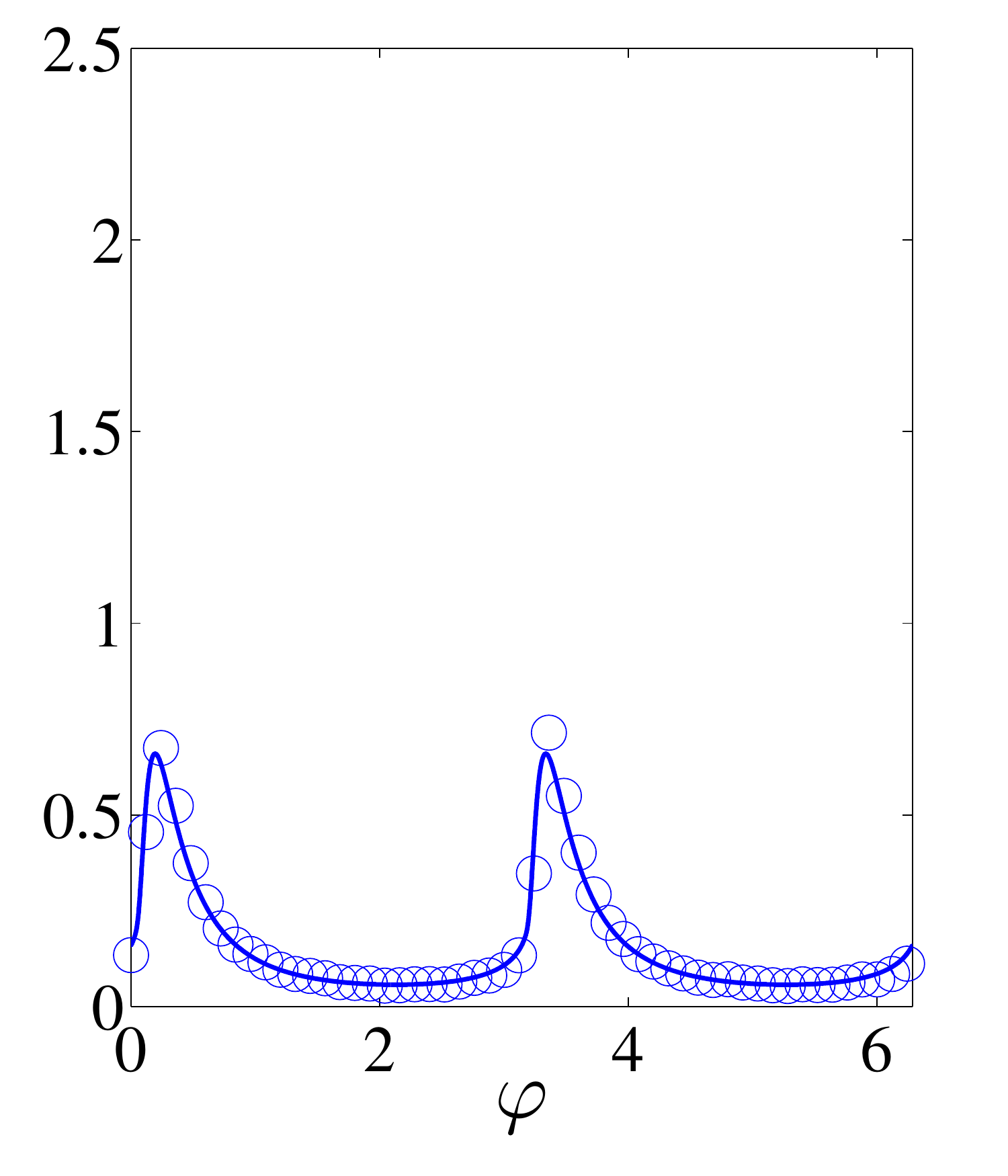}}\subfigure[\hspace{6cm}]{\includegraphics[scale=.35]{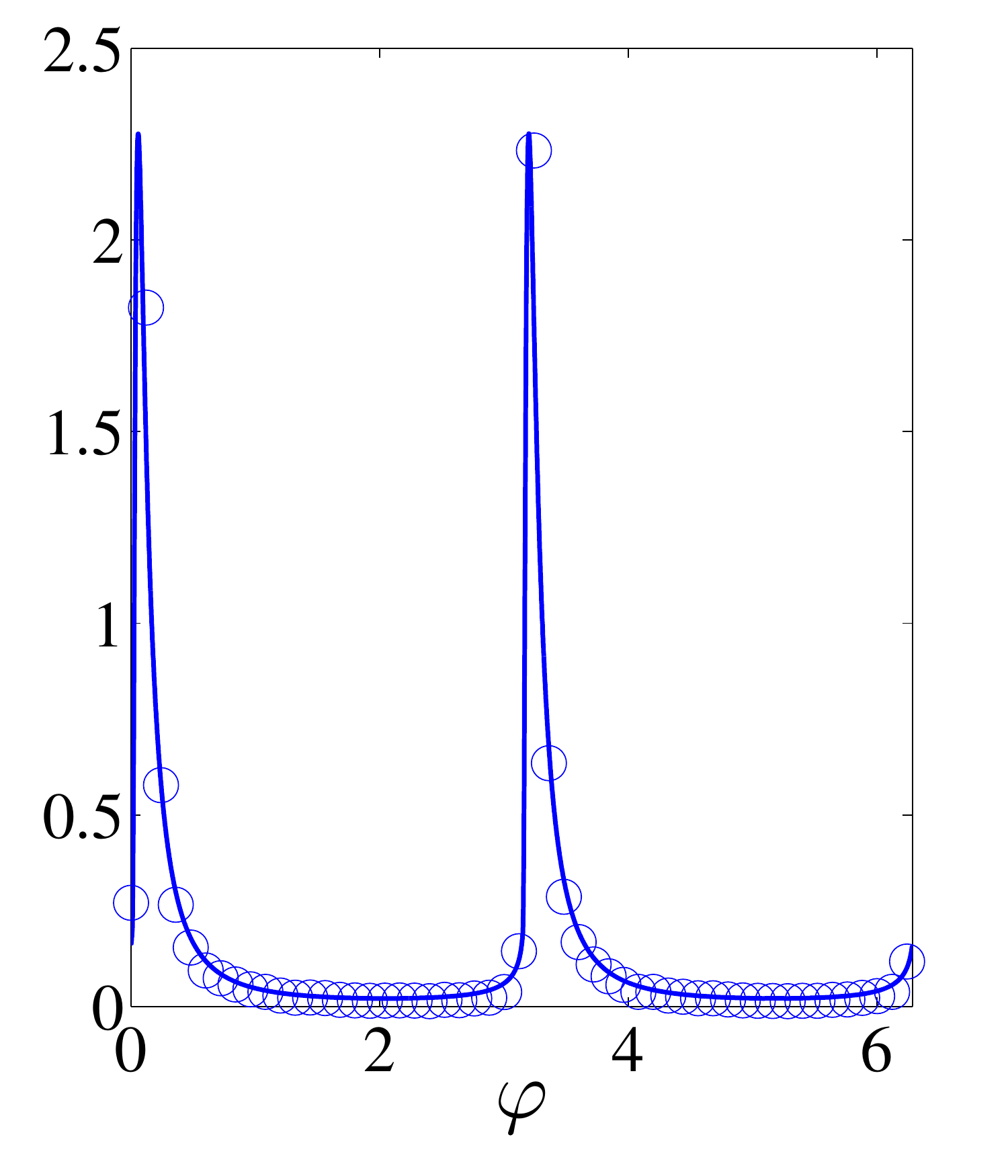}}
\end{center}
\caption{{\bf Approximations of the CR stationary density.}  The asymptotic approximation \eqref{eq:CRdensityapprox} (solid line) is plotted against Monte-Carlo simulations of the full system \eqref{eq:rphi} (open circles) for (a) $K_{CR}=1$, (b) $K_{CR}=2$, and (c) $K_{CR}=4$. In all figures, $\gamma=50$, $c=20$, and $\omega=1$. }\label{fig:CRdensityapprox}
\end{figure}

Starting from the Fickian form of the FP equation \eqref{eq:phaseFPFick}, we integrate once and expand terms in powers of $\varepsilon$ to obtain the leading order equation

\begin{equation*}
-J^0 = D^{CR,0}_{\varphi}(\varphi)\partial_{\varphi}u^0_{ss}-V^{CR,0}_{\varphi}(\varphi)u^0_{ss},
\end{equation*}
with $u_{ss}^0(0)=u_{ss}^0(2\pi/\omega)$ and $J^0= (2\pi/\omega)^{-1}$ is the leading order approximation to the flux. Owing to the simple forms of $V^{CR,0}_{\varphi}$ and $D^{CR,0}_{\varphi}$, we integrate the above equation and apply the boundary conditions to obtain

\begin{equation}
\label{eq:CRdensityapprox}
u^0_{ss}=\frac{1} {2\pi/\omega}\frac{i\omega} {K_{CR}^2} e^{-\Phi(\varphi)} \left[e^{A i} \textrm{Ei}(-A\cot(\omega \varphi)+Ai) - e^{-A i} \textrm{Ei}(-A\cot(\omega \varphi)-Ai)\right],
\end{equation}
where $A=2\omega/K_{CR}^2$, 

\begin{equation*}
\Phi(\varphi)\equiv\int \frac{-V^{CR,0}_{\varphi}(\tilde{\varphi})} {D^{CR,0}_{\varphi}(\tilde{\varphi})} d\tilde{\varphi}=\ln(\sin^2(\omega \varphi)) + A \cot(\omega \varphi),
\end{equation*}
and $\textrm{Ei}(x)$ is the exponential integral. Figure \ref{fig:CRdensityapprox} plots approximation $u_{ss}^0$ \eqref{eq:CRdensityapprox} against Monte-Carlo simulations of the full model \eqref{eq:rphi} for three different values of $K_{CR}$. Notice that bistability increases as $K_{CR}$ is increased and that our approximation is in excellent agreement with the simulations for all values of $K_{CR}$. Thus, bistability occurs when the product of $c\sigma$ is large enough, illustrating that the noise interacts with the vector field in a nonlinear way in order to produce the bistable switching behavior.

\begin{figure}[h!]
\begin{center}
{\includegraphics[scale=.6]{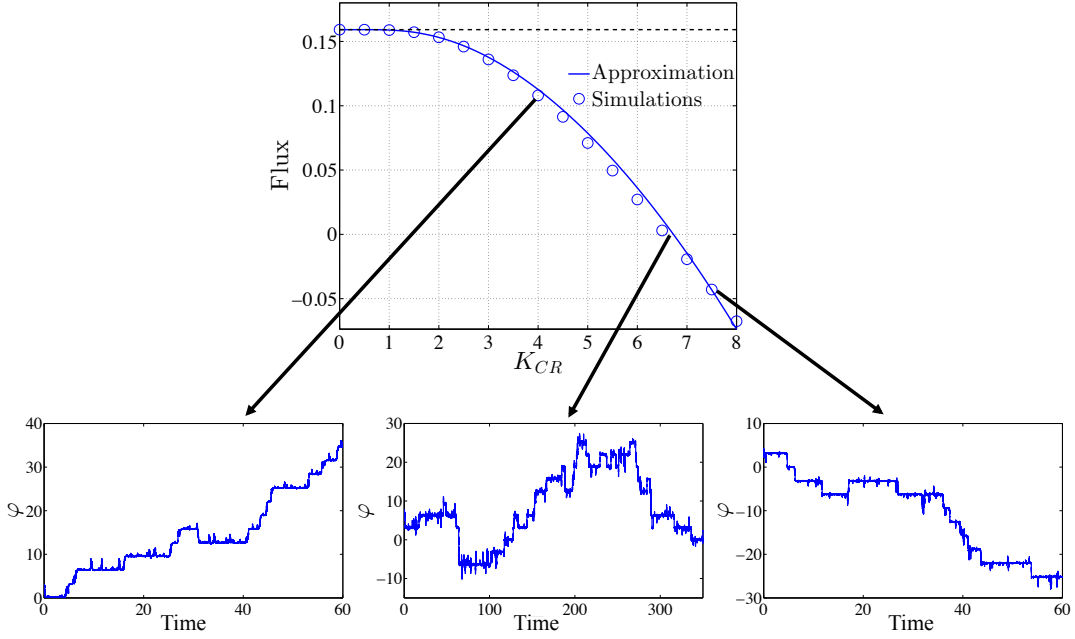}}
\end{center}
\vspace{-3cm}
\caption{{\bf Approximation of the CR flux.}  The asymptotic approximation \eqref{eq:CRfluxapprox} (solid line) is plotted as a function of $K_{CR}$ along with simulations of the full system \eqref{eq:rphi} open circles.  The dashed line is $(2\pi/\omega)^{-1}$ which is the frequency of the deterministic limit cycle and the zeroeth order approximation to the flux. Lower figures are sample phase trajectories of the full system \eqref{eq:rphi}.  From left to right $K_{CR}=4$, $K_{CR}=6.6$, and $K_{CR}=7.5$. In all figures, $\gamma=50$, $c=20$, and $\omega=1$.}\label{fig:CRfluxvK}
\end{figure}

Now that we have an approximation to the stationary density, we can compute our approximation to the stationary flux \eqref{eq:CRfluxapprox}.  Unfortunately, the integral in equation \eqref{eq:CRfluxapprox} can not be computed analytically, and we therefore approximate it using quadrature.  Figure \ref{fig:CRfluxvK} plots the approximation to the flux (solid line) along with the results from Monte-Carlo simulations of the full system \eqref{eq:rphi}.  The dashed line is $(2\pi/\omega)^{-1}$ which is the leading order approximation to the flux.  The lower panels plot example phase trajectories for three different values of $K_{CR}$.  To compute the flux numerically, we took advantage of the fact that the system is ergodic and simulated \eqref{eq:rphi} for a long time $T$.  Using the time series $\varphi(t)$ obtained from the simulation, we then computed the time average of the Ito drift 

\begin{equation*}
J=\frac{1} {2\pi/\omega} \frac{1} {T}\int_0^TI_{\varphi}^{CR}(\varphi(t))dt.
\end{equation*}
It can clearly be seen that the simulations are in excellent agreement with our approximation.  As $K_{CR}$ is increased, the flux decreases and eventually becomes negative.  The value of $K_{CR}$ where the flux becomes negative is also the value of $K_{CR}$ where the Ito drift starts to become negative.  Ignoring the $\varepsilon^2$ terms in $I_{\varphi}^{CR}$ \eqref{eq:CRexpansions}, we find that the value of $K_{CR}$ where the Ito drift first becomes negative is given by\footnote{Set $\min_{\varphi} I_{\varphi}^{CR,0}(\varphi)+I_{\varphi}^{CR,1}(\varphi)=1-\varepsilon\frac{K_{CR}^2} {2\omega} = 0$ and solve for $K_{CR}$.} $K_{CR}=\sqrt{2\omega/\varepsilon}$ which is approximately $6.32$ for the parameters in Figure \ref{fig:CRfluxvK}.  Thus, we can see that around this point, the flux is close to zero and eventually becomes negative as $K_{CR}$ is increased further. 

When the flux is positive, the phase jumps in the direction of the deterministic limit cycle (see bottom left panel in Figure \ref{fig:CRfluxvK}). This explains the behavior we saw in Figure \ref{fig:bistablephase} (a) as $K_{CR}=2$ for the parameters used. However, when the flux is close to zero, the phase still jumps fairly regularly, however the jumps appear to be equally likely in both the postive and negative directions (bottom middle panel in Figure \ref{fig:CRfluxvK}). Lastly, when the flux becomes negative, the process jumps in the direction opposite of the motion of the deterministic limit cycle (bottom right panel in Figure \ref{fig:CRfluxvK}).  Thus, it is clear that for the CR system, the flux determines the average direction of the phase jumps, but does not influence the rate at which the process jumps.


\subsection{AR System}

\begin{figure}[h!]
\begin{center}
\subfigure[\hspace{6cm}]{\includegraphics[scale=.35]{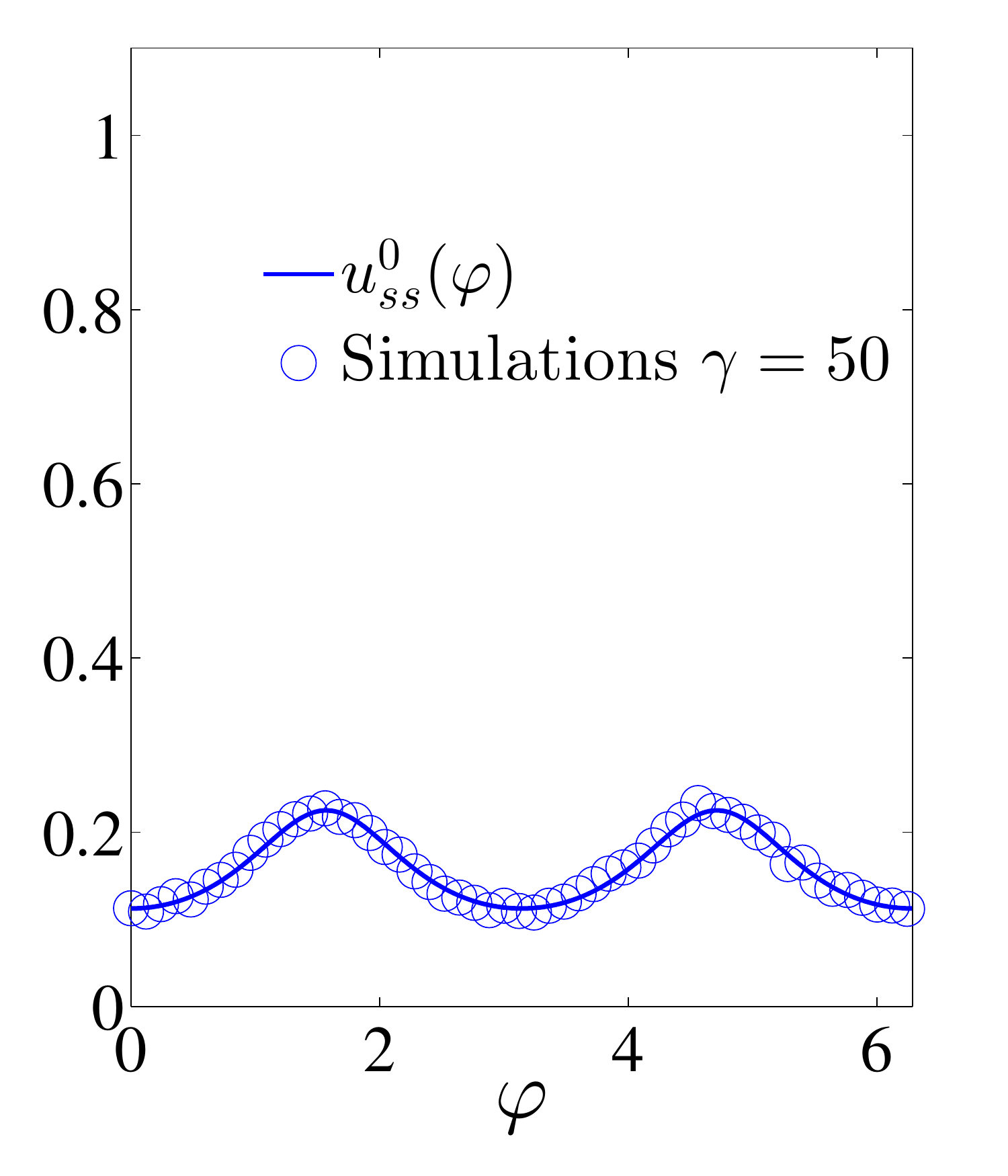}}\subfigure[\hspace{6cm}]{\includegraphics[scale=.35]{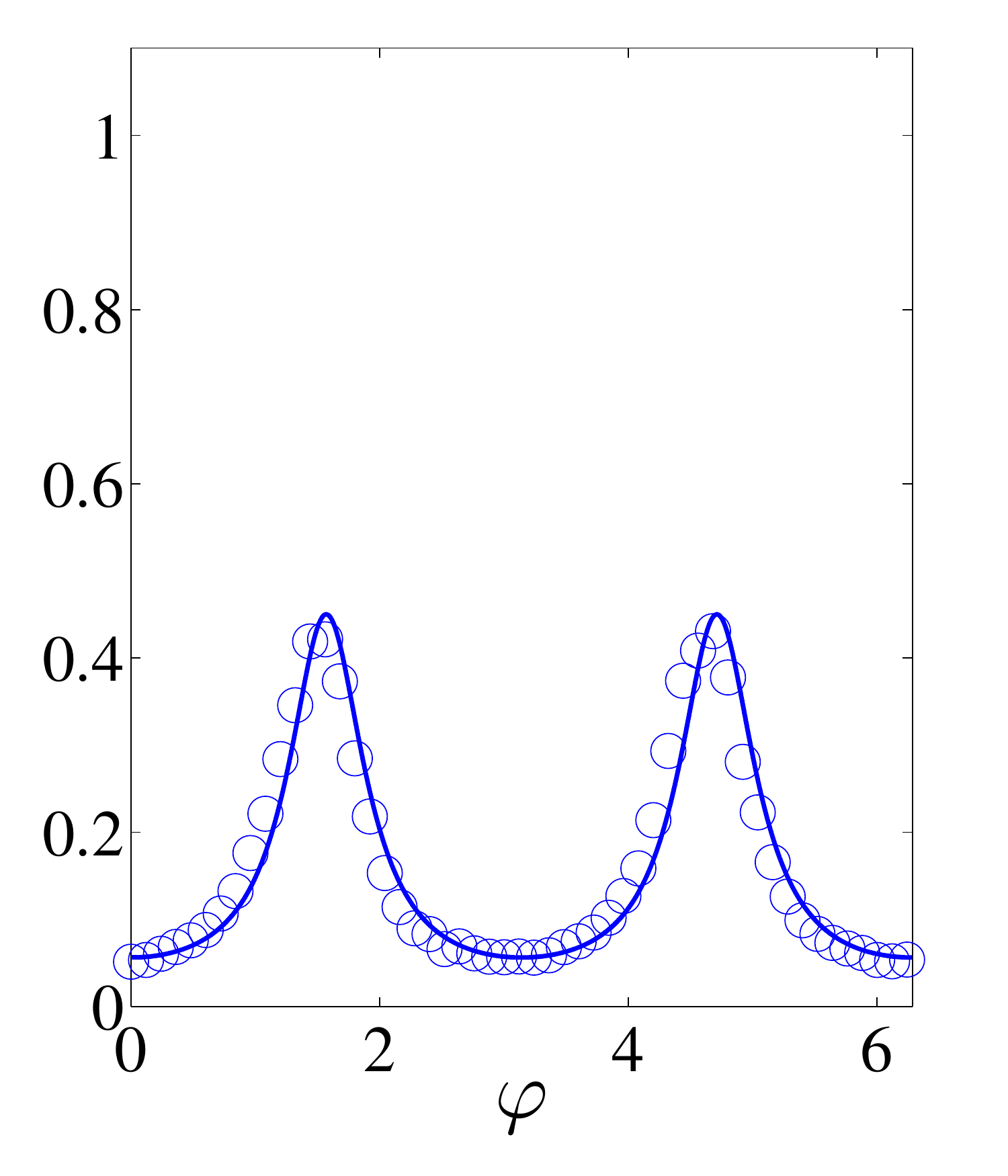}}\subfigure[\hspace{6cm}]{\includegraphics[scale=.35]{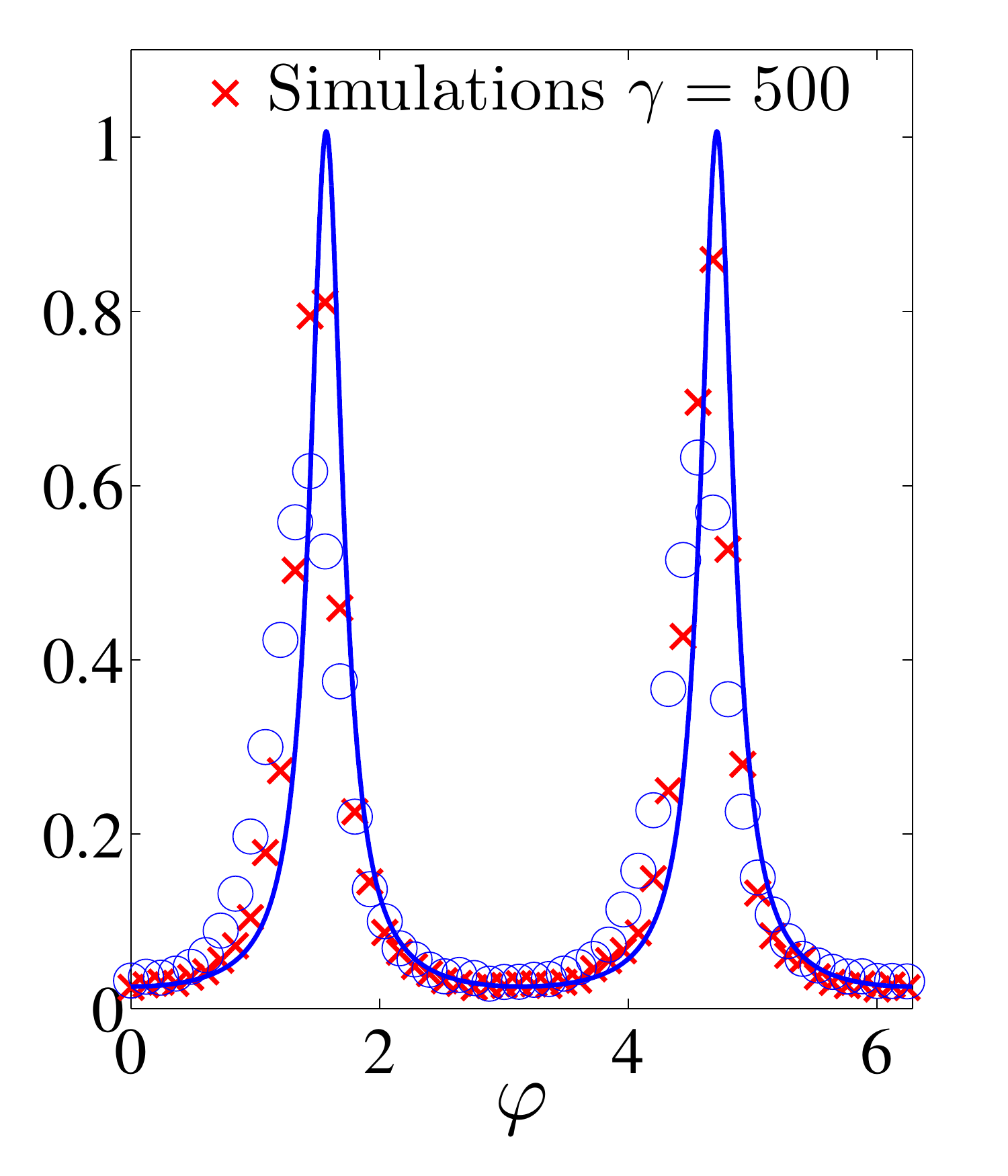}}
\end{center}
\caption{{\bf Approximations of the AR stationary density.}  The asymptotic approximation \eqref{eq:ARdensityapprox} (solid line) is plotted against Monte-Carlo simulations of the full system \eqref{eq:rphi} (open circles) for (a) $K_{CR}=2$, (b) $K_{CR}=3.5$, and (c) $K_{CR}=3.9$. In all figures, $\gamma=50$, $c=20$, and $\omega=1$.  The crosses in (c) are simulations of the full system with $\gamma=50$. }\label{fig:ARdensityapprox}
\end{figure}

As in the previous section, we introduce the parameters $\varepsilon=\frac{1} {c}$ and $K_{AR}=\frac{\sigma^2} {\varepsilon}$. Note the difference between $K_{CR}$ and $K_{AR}$.  Substituting $c=\frac{1} {\varepsilon}$ and $\sigma^2=\varepsilon K_{AR}$ into equation \eqref{eq:ARstuff}, we obtain

\begin{eqnarray}
\label{eq:ARexpansions}
I^{AR}_{\varphi}(\varphi) &=& 1-\frac{K_{AR}} {4}\sin^2(\omega\varphi)-\varepsilon\frac{K_{AR}} {2\omega}\sin(2\omega\varphi) \\
V^{AR}_{\varphi}(\varphi) &=& 1-\frac{K_{AR}} {4} \sin^2(\omega\varphi) \nonumber \\
D^{AR}_{\varphi}(\varphi) &=& \varepsilon\frac{K_{AR}} {2\omega} \cos^2(\omega\varphi)\nonumber.
\end{eqnarray}
With this scaling, it is very easy to see that $I^{AR}_{\varphi}\approx V^{AR}_{\varphi}$ and that $D^{AR}_{\varphi}\approx 0$.  Using equation \eqref{eq:phaseFPFick}, and noting that the leading order diffusivity is $0$, we obtain a very simple equation for the leading order stationary density $u_{ss}^0$

\begin{equation}
u_{ss}^0(\varphi)=\frac{J_0} {V^{AR}_{\varphi}(\varphi)}\label{eq:ARdensityapprox}.
\end{equation}
From the above equation, we can see that the leading order flux is the same as the normalization constant for the density.  Enforcing that $\int_0^{2\pi/\omega}u_{ss}^0(\varphi)d\varphi=1$ we find

\begin{equation}
J_0=\frac{\sqrt{4-K_{AR}}} {4\pi/\omega}\label{eq:ARfluxapprox}.
\end{equation}
However, notice that this approximation is only valid for $K_{AR}<4$.  This is an interesting point that we will return to in the next section.  Figure \ref{fig:ARdensityapprox}  plots our approximation $u_{ss}^0$ against Monte-Carlo simulations of the full model \eqref{eq:rphi} for three different values of $K_{AR}$. Again, notice that bistability increases as $K_{AR}$ is increased.  However, notice that our approximation begins to break down near $K_{AR}=4$ (Fig. \ref{fig:ARdensityapprox} (c)), but this can be alleviated by taking $\gamma$ to be an order of magnitude larger (compare circles to crosses).  This change in the behavior of the system can be seen even more clearly when we plot the flux as a function of $K_{AR}$ as in Figure \ref{fig:ARdensityapprox}.  In contrast to the CR system, the flux in the AR system quickly goes to zero as $K_{AR}$ is increased.  We can see that the analytical approximation works very well for values of $K_{AR}$ away from $4$ (compare circles to solid line). Taken together with Figure \ref{fig:ARdensityapprox} (c), this shows that, near $K_{AR}=4$, the $\mathcal{O}(1/\gamma)$ terms that we are ignoring in the phase reduction \eqref{eq:phase} become significant.  After $K_{AR}=4$, our approximation \eqref{eq:ARdensityapprox} is no longer valid. However, numerical simulations show that the flux is very close to zero.

\begin{figure}[h!]
\begin{center}
{\includegraphics[scale=.6]{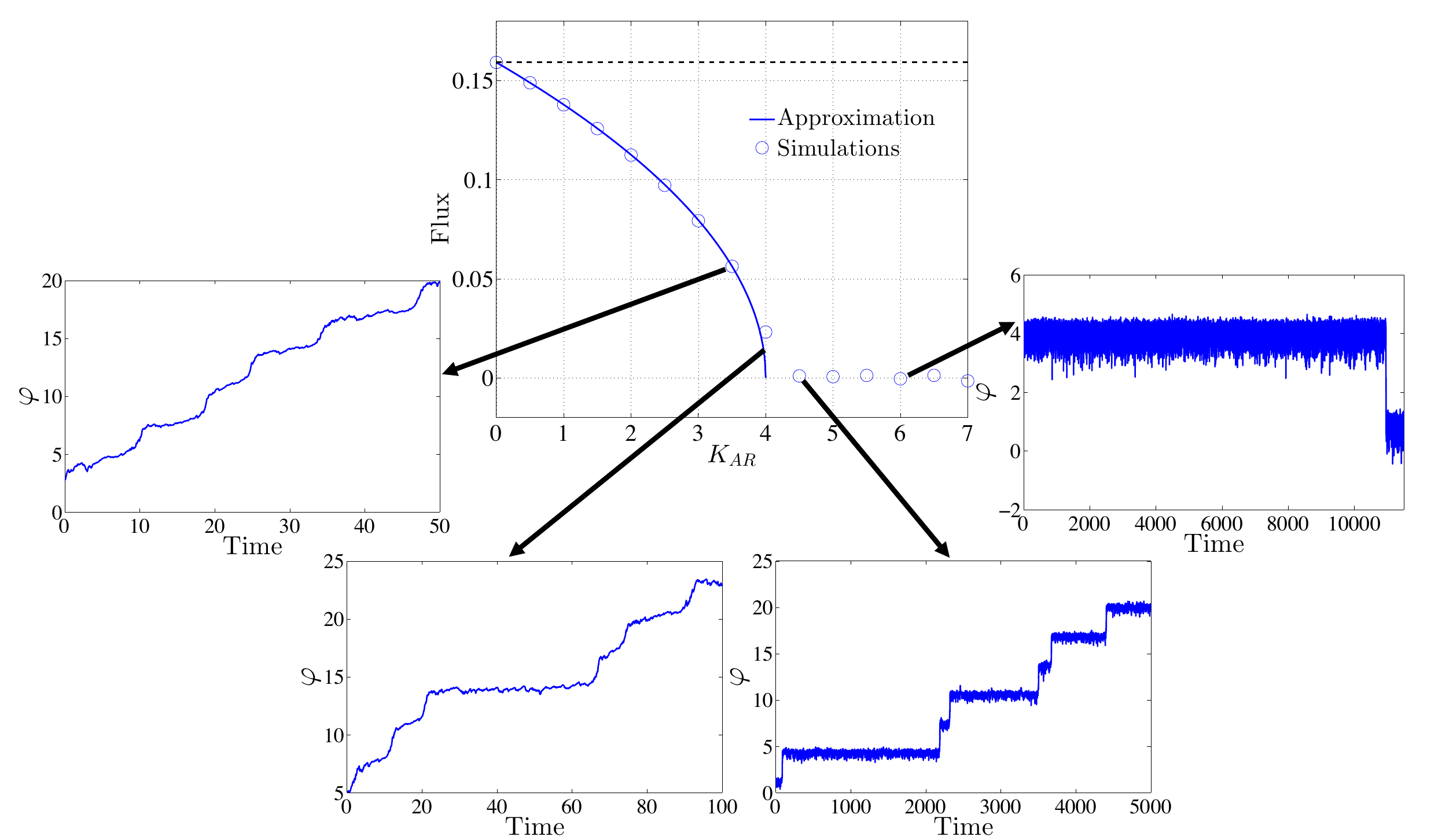}}
\end{center}
\vspace{0cm}
\caption{{\bf Approximation of the AR flux.}  The asymptotic approximation \eqref{eq:ARfluxapprox} (solid line) is plotted as a function of $K_{AR}$ along with simulations of the full system \eqref{eq:rphi} open circles.  The dashed line is $(2\pi/\omega)^{-1}$ which is the frequency of the deterministic limit cycle. Lower figures are sample phase trajectories of the full system \eqref{eq:rphi}.  From left to right $K_{AR}=3.5$, $K_{AR}=3.9$, $K_{AR}=4.5$, and $K_{AR}=6$. In all figures, $\gamma=50$, $c=20$, and $\omega=1$.}\label{fig:ARfluxvK}
\end{figure}

As with the CR system, we can see that the flux is a good indicator of the average direction of the phase jumps (see the bottom panels in Fig. \ref{fig:ARfluxvK}). However, it is difficult to interpret the results in the $K_{AR}>4$ regime as the flux is very close to zero.  In contrast to the CR system, the rate of switching in the AR system also decreases with $K_{AR}$. Indeed, at $K_{AR}=3.5$, the system displays frequent phase jumps in the positive direction, but at $K_{AR}=6$ the system displays incredibly rare jumps in the negative direction (compare x-axis scales in lower left and lower right panels) which is similar to the behavior we saw in Figure \ref{fig:bistablephase} (b) as $K_{AR}=6$ for the parameters used.  However, this decrease in the jump rate cannot be due to the flux, as we saw that with near zero flux in the CR system, the rate of switching was still fairly frequent. In the next section, we seek to explain why there are differences in the switching frequency.


\section{Switching Frequency is Determined by Diffusivity}

We have shown that the onset of bistable switching in the CR and AR systems occurs when the Fickian drift becomes negative, and that the direction of the switching (either in the same or opposite direction of the deterministic limit cycle) is indicated by the sign of the stationary probability flux. Here we will show that the frequency of the switching events can be explained by the relative strengths of the Ito drifts and diffusivities in the reduced phase system \eqref{eq:phase}.


\subsection{Rare Switching in AR System is Caused by Weak Diffusivity}
 
\begin{figure}[h!]
\begin{center}
\subfigure[\hspace{11cm}]{\includegraphics[scale=.2]{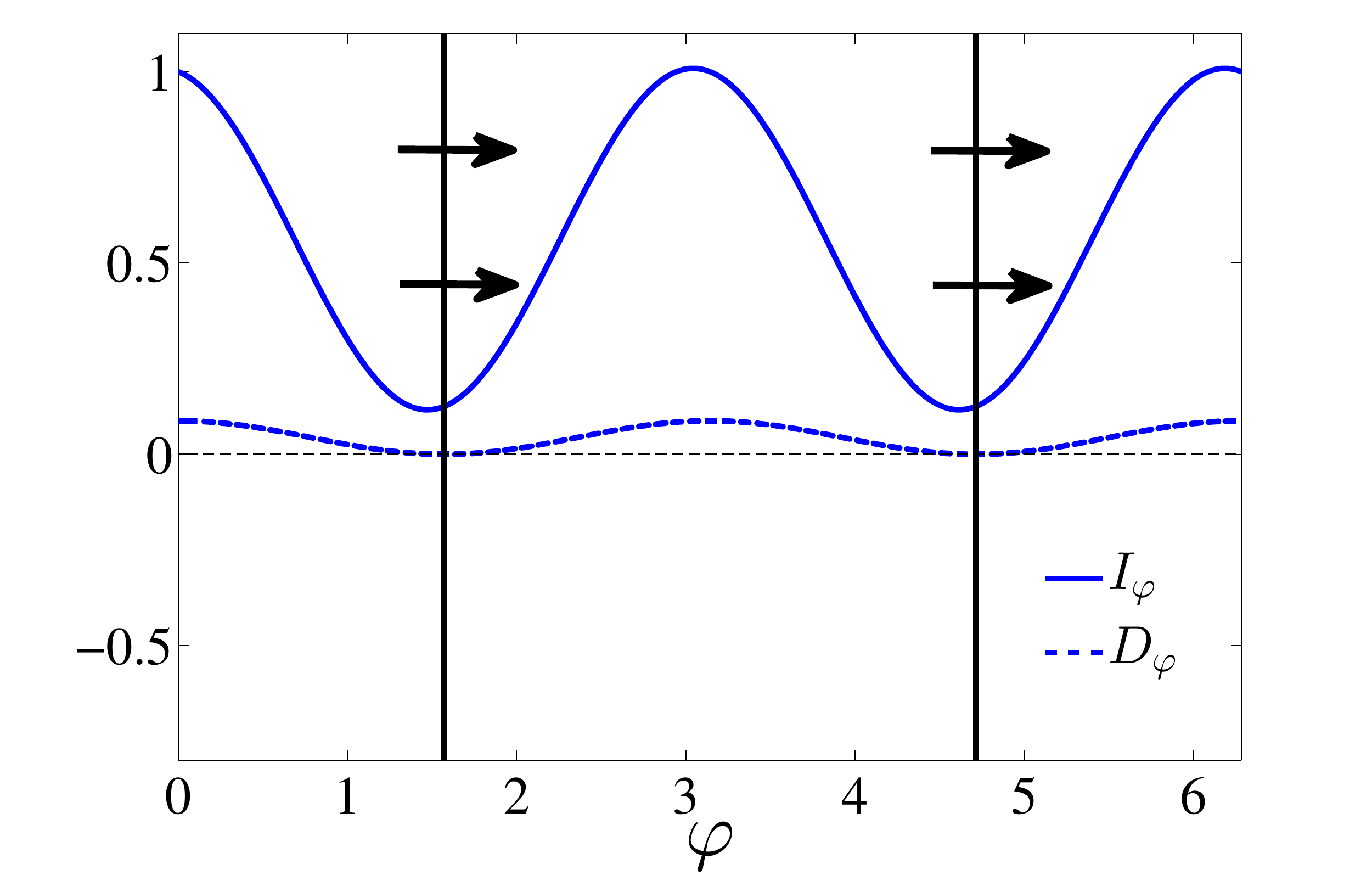}\includegraphics[scale=.2]{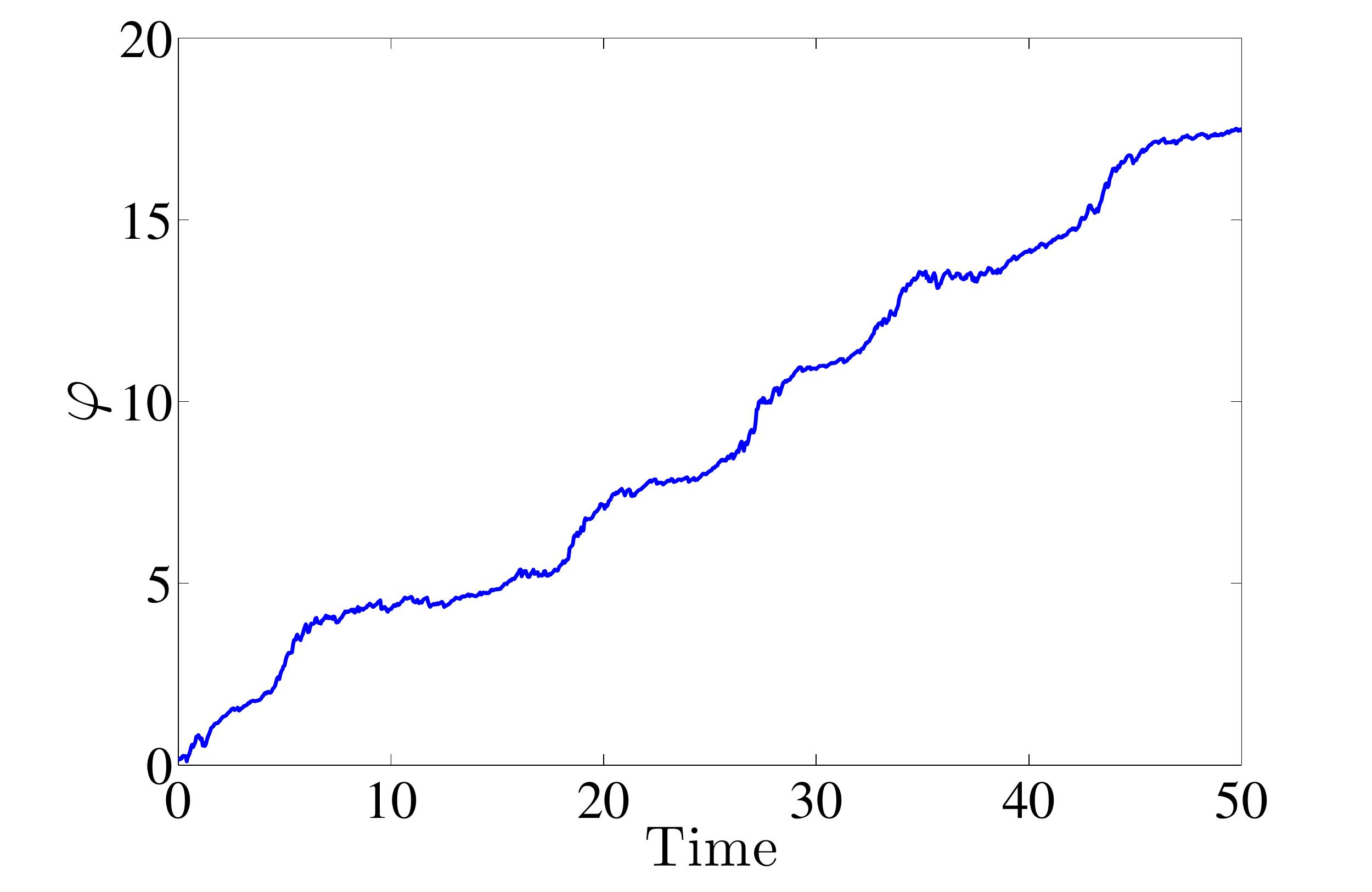}}\\
\subfigure[\hspace{11cm}]{\includegraphics[scale=.2]{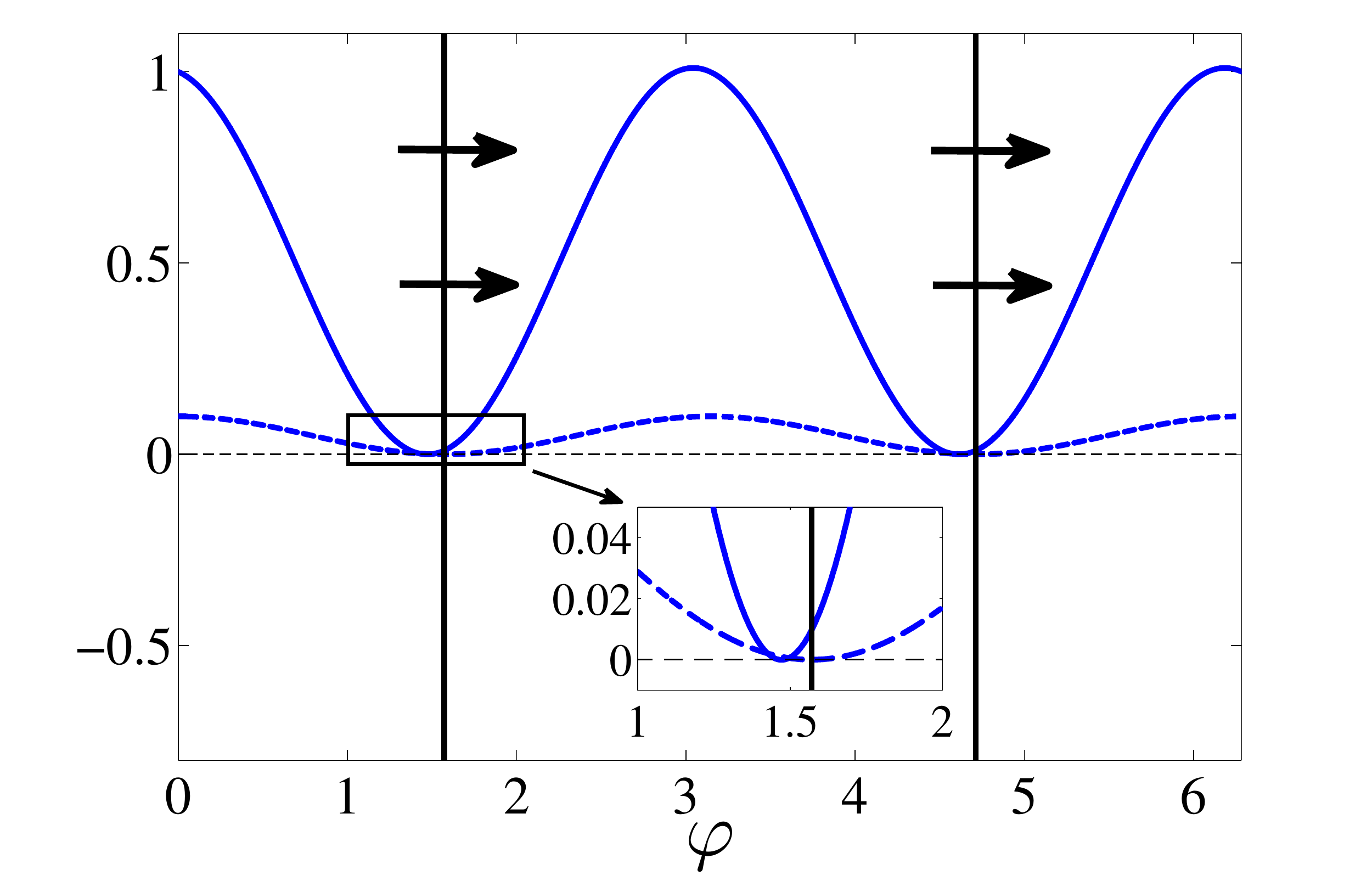}\includegraphics[scale=.2]{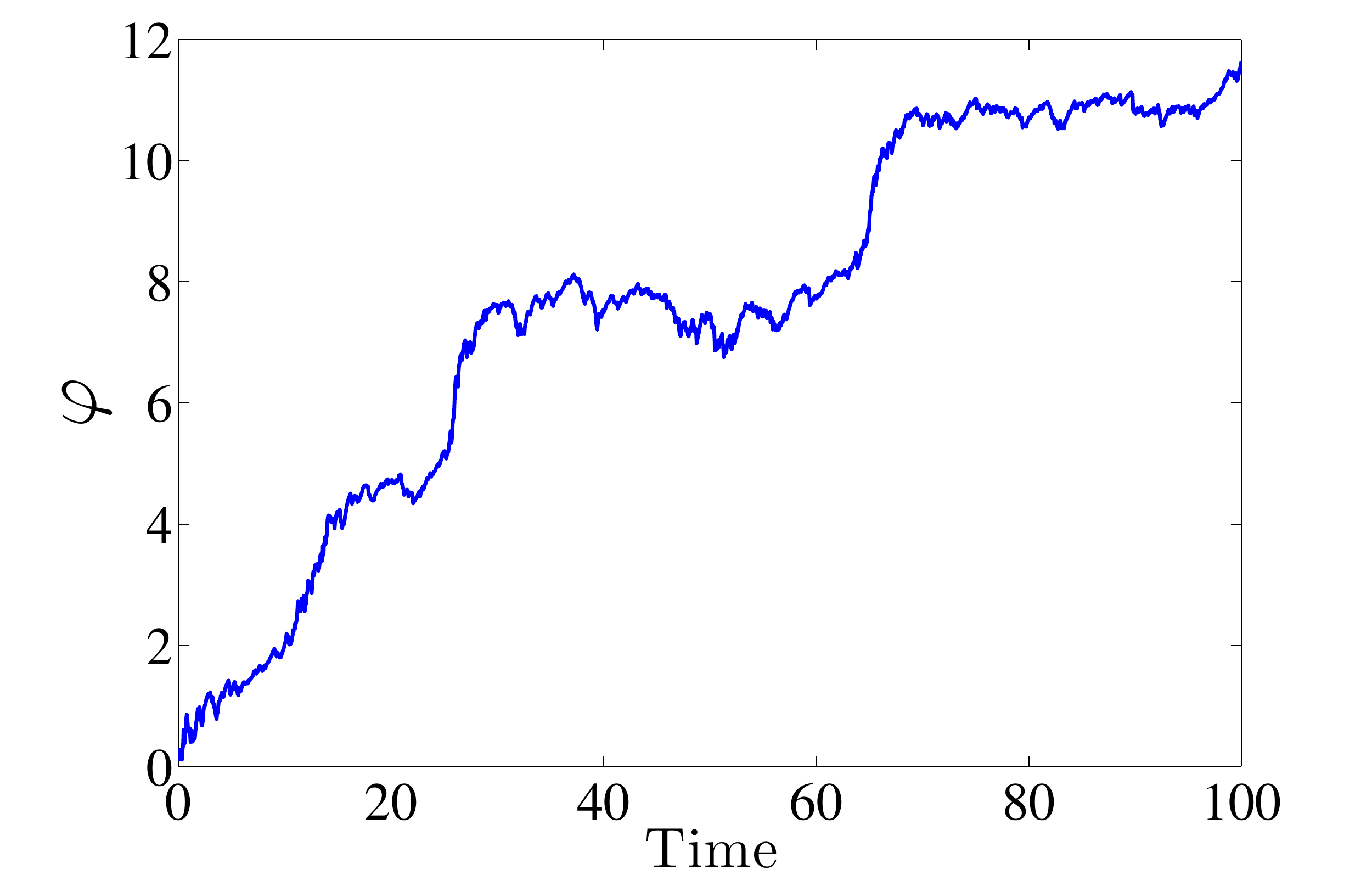}}\\
\subfigure[\hspace{11cm}]{\includegraphics[scale=.2]{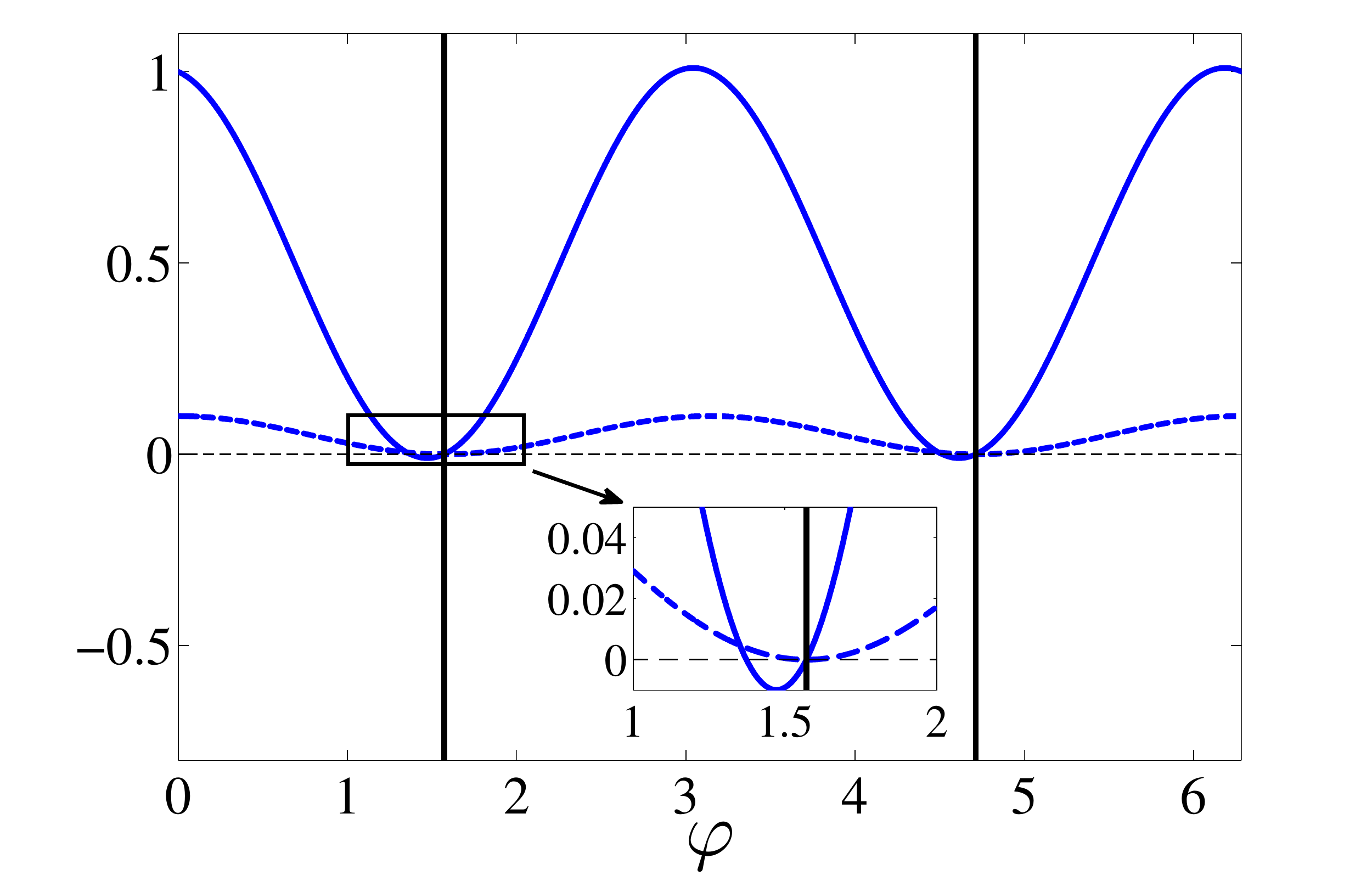}\includegraphics[scale=.2]{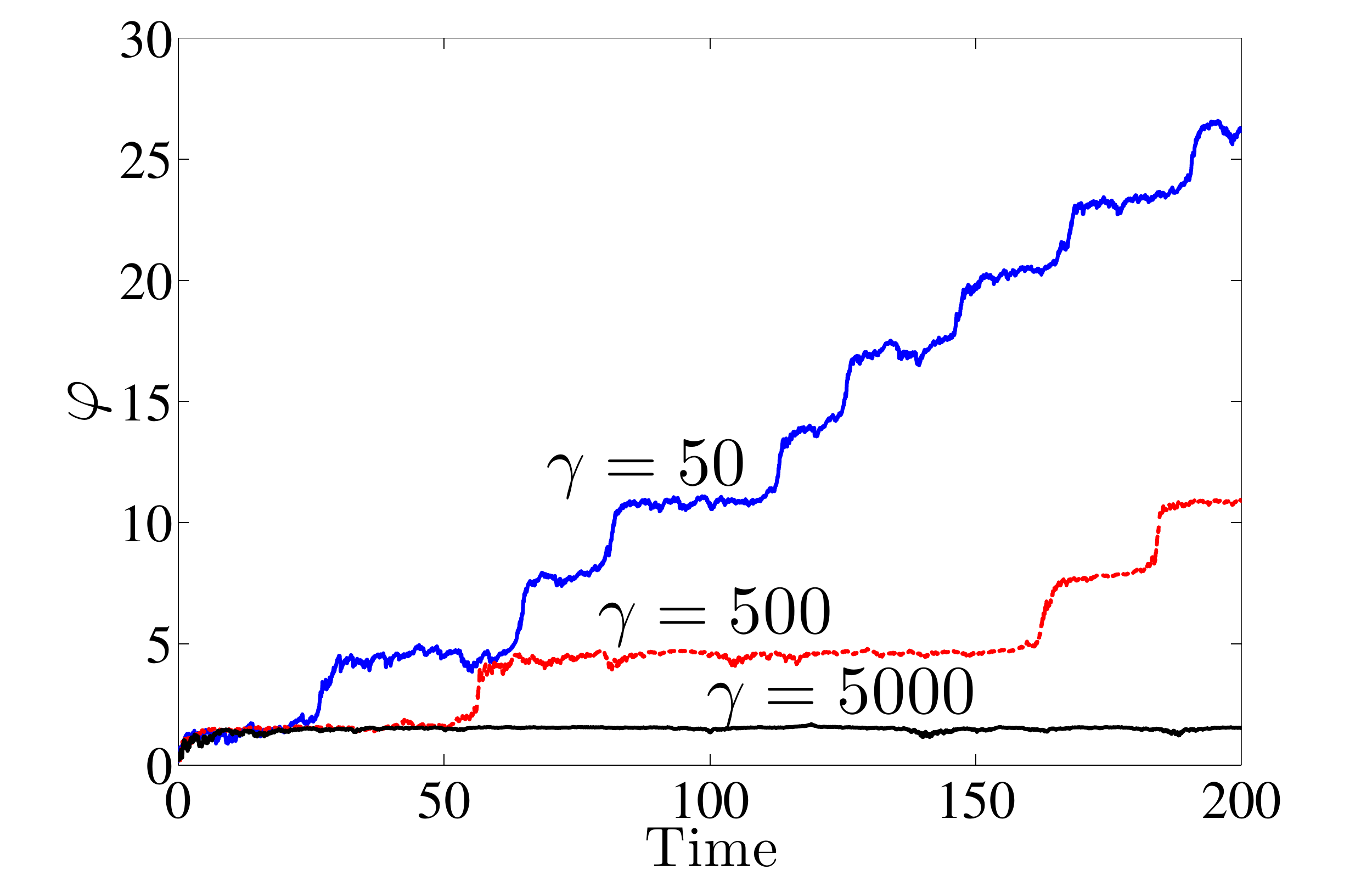}}\\
\subfigure[\hspace{11cm}]{\includegraphics[scale=.2]{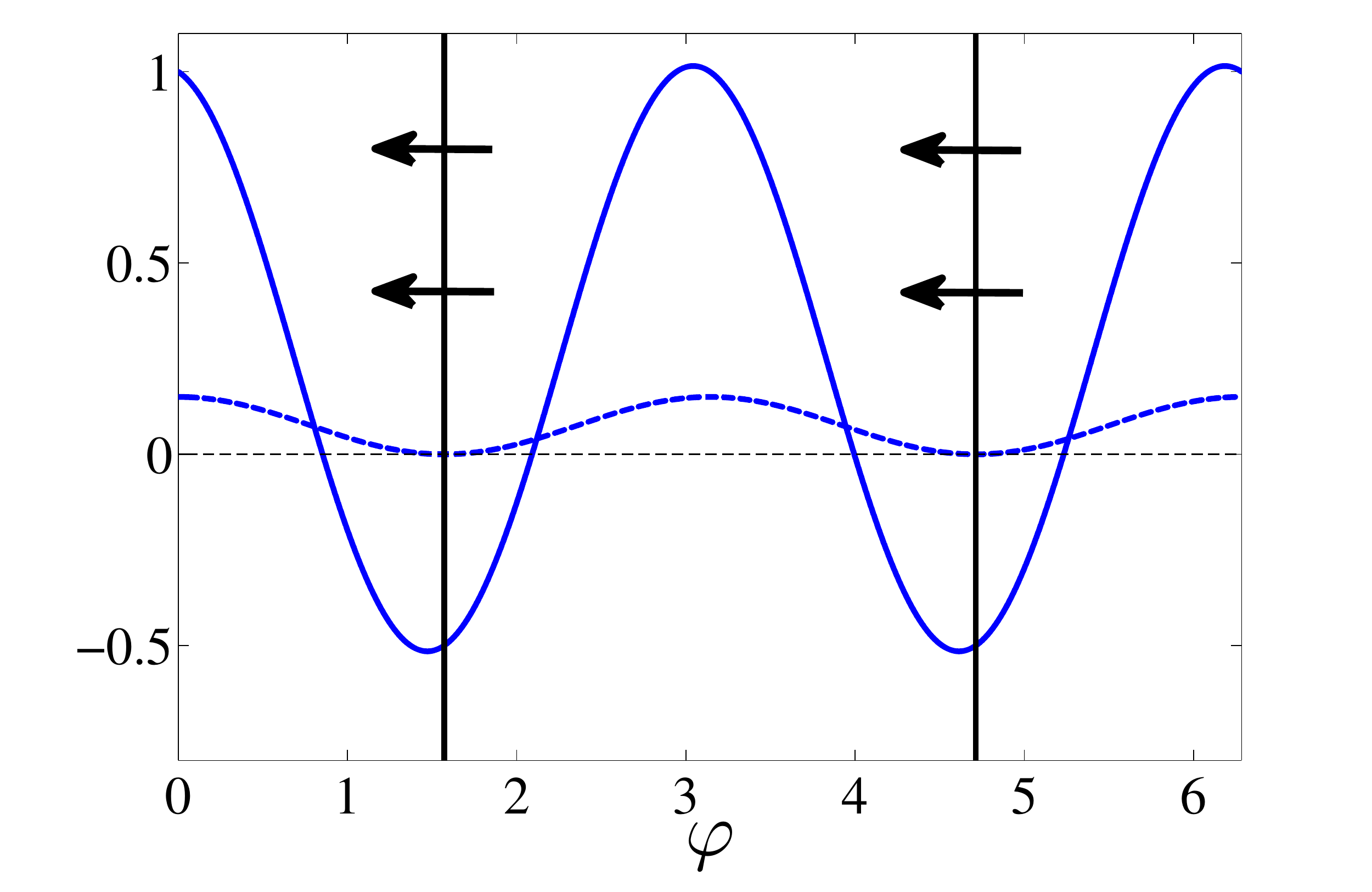}\includegraphics[scale=.2]{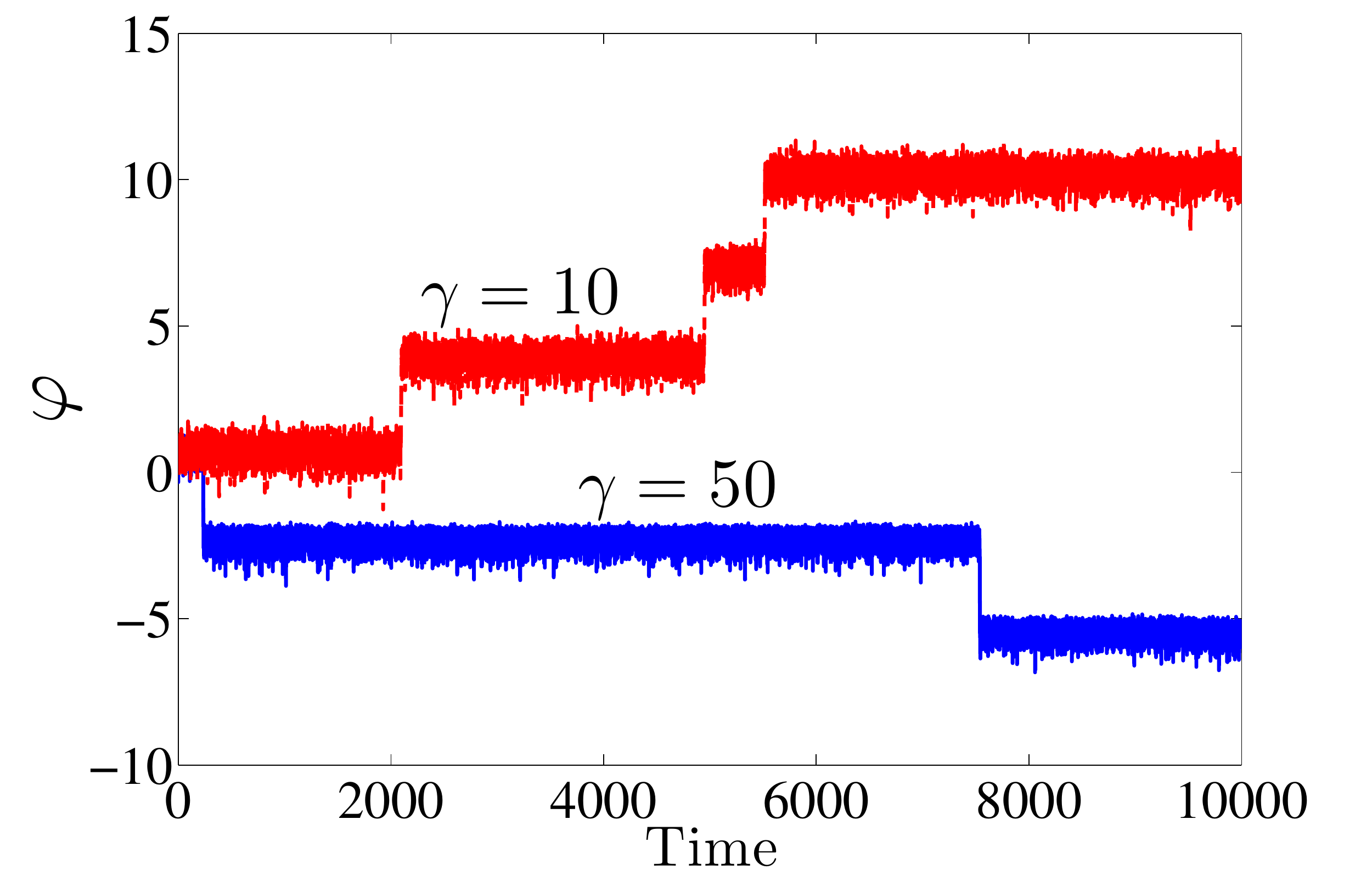}}
\end{center}
\caption{{\bf AR system is drift dominated.}  Left panels show the Ito drift (solid lines) and diffusivity (dashed lines) of the phase reduced system \eqref{eq:phase}.  Solid vertical lines indicate the ratchet points (where the diffusivity is zero) and the horizontal arrows indicate the sign of the Ito drift at the ratchet points.  Right panels show sample phase trajectories of the full system \eqref{eq:rphi}. For all plots, $c=20$, $\omega=1$, and unless indicated otherwise $\gamma=50$. (a) $K_{AR}=3.5$, (b) $K_{AR}=3.961$, (c) $K_{AR}=4$, and (d) $K_{AR}=6$. }\label{fig:ARItodiff}
\end{figure}

To understand both the decreasing rate of switching and the change in switching direction of the AR system, we examine the contributions of the Ito drift and diffusivity to the reduced phase model using the scalings introduced in the previous section.  First, recall the diffusivity for the AR model given in \eqref{eq:ARexpansions}.  Notice two things about the AR diffusivity: (1) that it is order $\varepsilon$; and (2) that it has two zeros $\varphi^*=\frac{\pi} {2\omega},\frac{3\pi} {2\omega}$ regardless of the values of $K_{AR}$ and $\varepsilon$.  We will refer to these zeros of the diffusivity as ``ratchet points''.  The only way that the reduced phase system can cross these ratchet point is if the Ito drift is nonzero at these points.  Figure \ref{fig:ARItodiff} (a) plots the Ito drift (solid) and diffusivity (dashed) as a function of $\varphi$ in the left panel and a sample stochastic phase trajectory of the full system \eqref{eq:rphi} in the right panel when $K_{AR}=3.5$.  It is clear that the Ito drift is always significantly larger than the diffusivity, and thus is the main determinant of the motion of the stochastic process.  This can be seen in the right panel of Fig. \ref{fig:ARItodiff} (a) where the phase variable increases steadily over time.  The effect of the changing Ito drift can be seen in the slope of the trajectory as the phase variable increases more slowly when it is near the two ratchet points.   In this case, the Ito drift is positive at the ratchet points (as indicated by the horizontal arrows) which indicates that the stochastic trajectory will pass over the ratchet points in the direction of the deterministic limit cycle.  

As $K_{AR}$ is increased to $3.961$ as in Fig. \ref{fig:ARItodiff} (b), the Ito drift near the ratchet points becomes the same order at the diffusivity (inset, left panel). Thus, the process begins to have a more difficult time passing over the ratchet points, and begins to display the phase jumps seen previously (right panel). The jumps occur in the positive direction as the Ito drift at the ratchet point is still positive (horizontal arrows, left panel).  When $K_{AR}=4$ (Fig. \ref{fig:ARItodiff} (c)), an interesting phenomenon occurs: the Ito drift is exactly zero at the ratchet points (to see this, plug $\varphi^*=\frac{\pi} {2\omega},\frac{3\pi} {2\omega}$ into the equation for $I^{AR}_{\varphi}$ in \eqref{eq:ARexpansions}).  Thus, the phase reduced system \eqref{eq:phase} predicts that once a trajectory arrives at a ratchet point, it will be stuck there indefinitely.  However, it is clear from the $\gamma=50$ sample trajectory in the right panel of Fig. \ref{fig:ARItodiff} (c) that the $\mathcal{O}(1/\gamma)$ terms in the phase reduction become significant as the trajectory of the full system \eqref{eq:rphi} is able to pass over the ratchet points. This discrepency between the full system and the phase reduced system is caused by the fact that the phase reduction ignores amplitude fluctuations. That is, for these values of $K_{AR}$ and $\gamma$, fluctuations in the amplitude variable allow the full system \eqref{eq:rphi} to move past the ratchet points. However, as $\gamma$ is increased, the trajectory has a more difficult time overcoming the ratchet points ($\gamma=500$ and $\gamma=5000$ traces).  Thus, as $\gamma\to \infty$ the stationary density for $\varphi$ converges to a delta function at one of the two ratchet points depending on the initial conditions.  

Lastly, when $K_{AR}>4$ (Fig. \ref{fig:ARItodiff} (d)), the Ito drift becomes negative at the ratchet points, and the phase reduction predicts that the process can only move past the ratchet points in the opposite direction of the deterministic limit cycle.  However, the Ito drift is only negative in a neighborhood of the ratchet points, and is strongly positive everywhere else.  Furthermore, the diffusivity is still much smaller than the Ito drift.  Thus, away from the ratchet points, the Ito drift dictates that the process move in the same direction of the limit cycle.  Therefore, the phase reduction predicts that the only way that the process can jump between ratchet points is via a rare event where diffusion overcomes the positive Ito drift and the process jumps in the opposite direction of the deterministic limit cycle (right panel, $\gamma=50$ trace).  Thus, the rare switching in the AR system is caused by the fact that the Ito drift is negative near the ratchet points, but positive \emph{and larger than the diffusivity} away from the ratchet points.  In contrast to the phase reduced system, if $\gamma$ is made smaller, fluctuations in the amplitude variable allow the full system \eqref{eq:rphi} to pass over the ratchet points in the same direction of the deterministic limit cycle (right panel, $\gamma=10$ trace).  Note also that the small diffusivity is the reason why our asymptotic approximation \eqref{eq:ARdensityapprox} breaks down when $K_{AR}>4$.  That is, our approximation assumes that the diffusivity is $0$ for all $\varphi$.  However, close to the points where the Ito drift first becomes negative, the diffusivity is non-negligible (and is in fact the reason why a switch can actually occur).

We can quantify the accuracy of the phase reduced system \eqref{eq:phase} by computing the $L^2$ norm of the difference between the stationary density obtained from solving the FP equation for the reduced system \eqref{eq:phaseFPIto} $u_{ss}$ and the stationary density obtained from Monte-Carlo simulation $u_{ss}^{MC}$.  More precisely, we take the error to be

\begin{equation}
\textrm{Error} = \left(\int_0^{2\pi/\omega}[u_{ss}(\varphi)-u_{ss}^{MC}(\varphi)]^2 d\varphi\right)^{1/2}.
\label{eq:error}
\end{equation}

\begin{figure}[h!]
\begin{center}
\includegraphics[scale=.35]{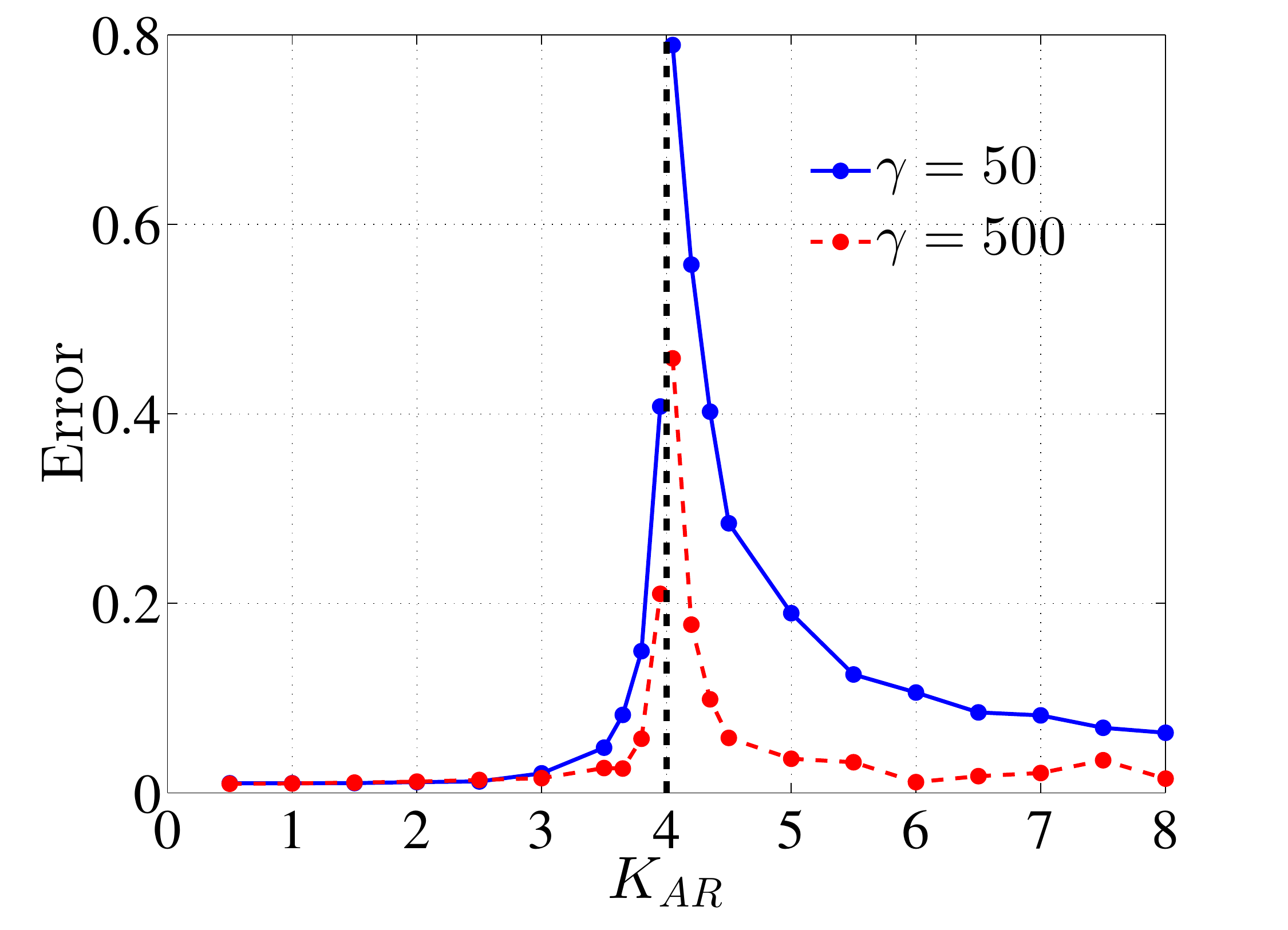}
\end{center}
\caption{{\bf Error in the phase model approximation of the stationary density for the AR system as a function of $K_{AR}$.} $L^2$ norm of the difference between the stationary density obtained from solving \eqref{eq:phaseFPIto} and the stationary density obtained from Monte-Carlo simulations of the full system \eqref{eq:rphi} as a function of $K_{AR}$ when $c=20$, $\omega=1$, $\gamma=50$ (solid line), and $\gamma=500$ (dashed line). }\label{fig:ARerror}
\end{figure}

Figure \ref{fig:ARerror} plots this error as a function of $K_{AR}$ and two different values for $\gamma$.  Note that to solve \eqref{eq:phaseFPIto}, we found it best to use a forward finite difference scheme when the flux was positive, and a backwards finite difference scheme when the flux was negative.  It is clear that the phase reduced model is highly accurate for $K_{AR}<4$, but the error quickly increases as $K_{AR}\to 4_{+}$.  Interestingly, the $\mathcal{O}(1/\gamma)$ terms appear to be more important for accuracy when $K_{AR}>4$.  Regardless, increasing $\gamma$ causes the error to decrease for all values $K_{AR}\neq 4$ (compare solid and dashed lines).

\subsection{Faster Switching in the CR System is Caused by Diffusivity Being Larger than the Ito Drift}

\begin{figure}[h!]
\begin{center}
\subfigure[\hspace{9cm}]{\includegraphics[scale=.2]{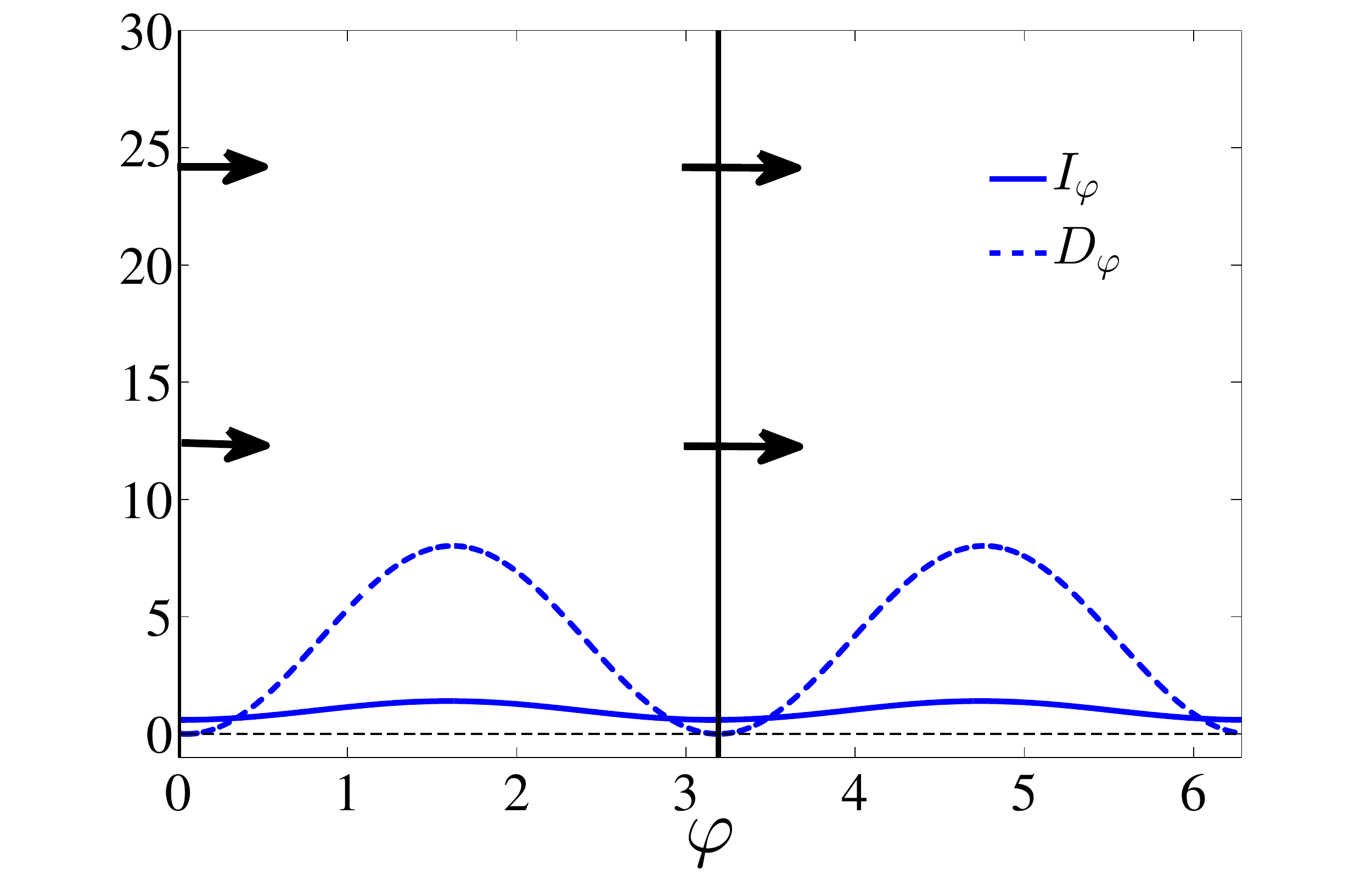}\includegraphics[scale=.2]{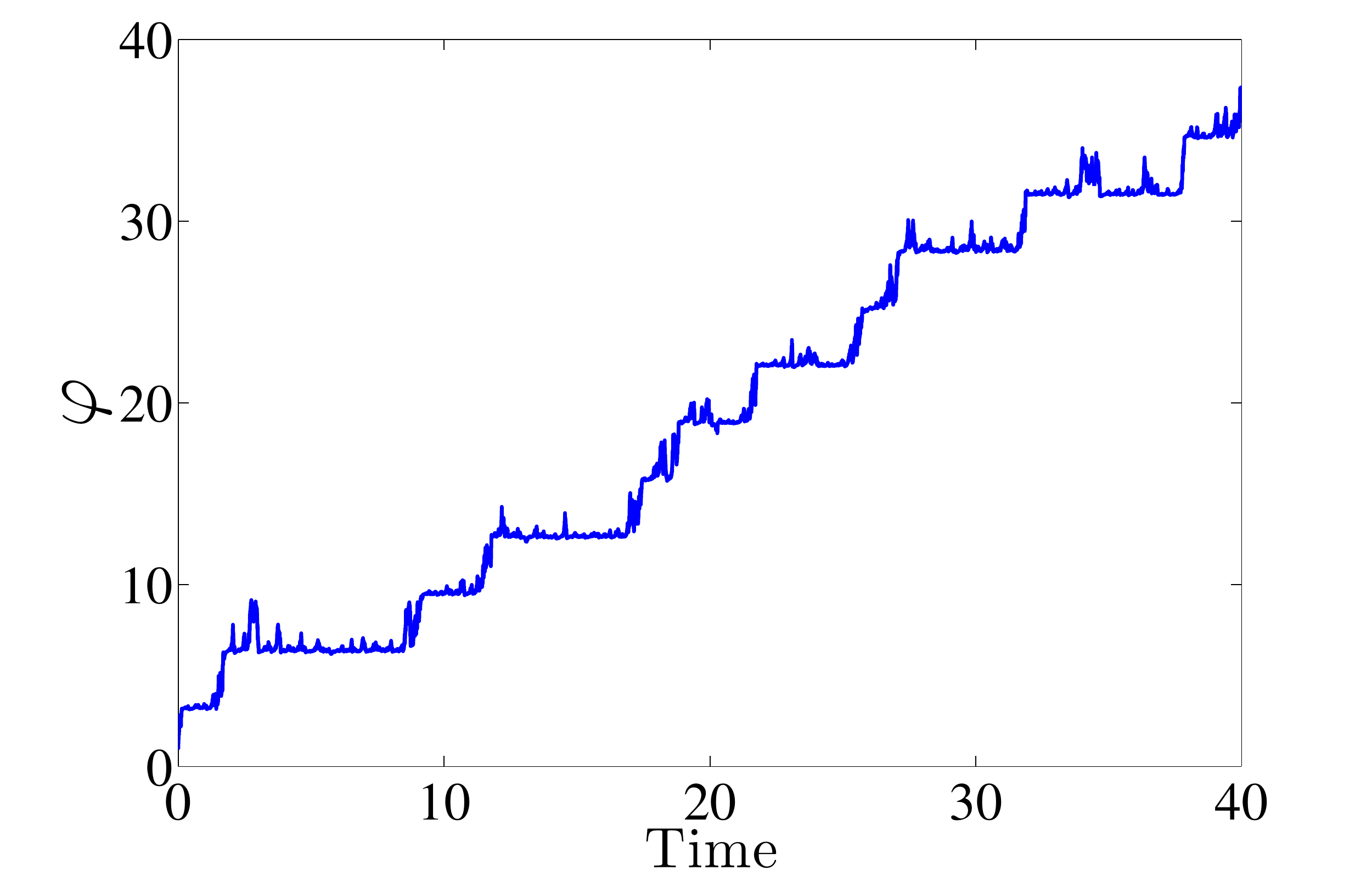}}\\
\subfigure[\hspace{9cm}]{\includegraphics[scale=.2]{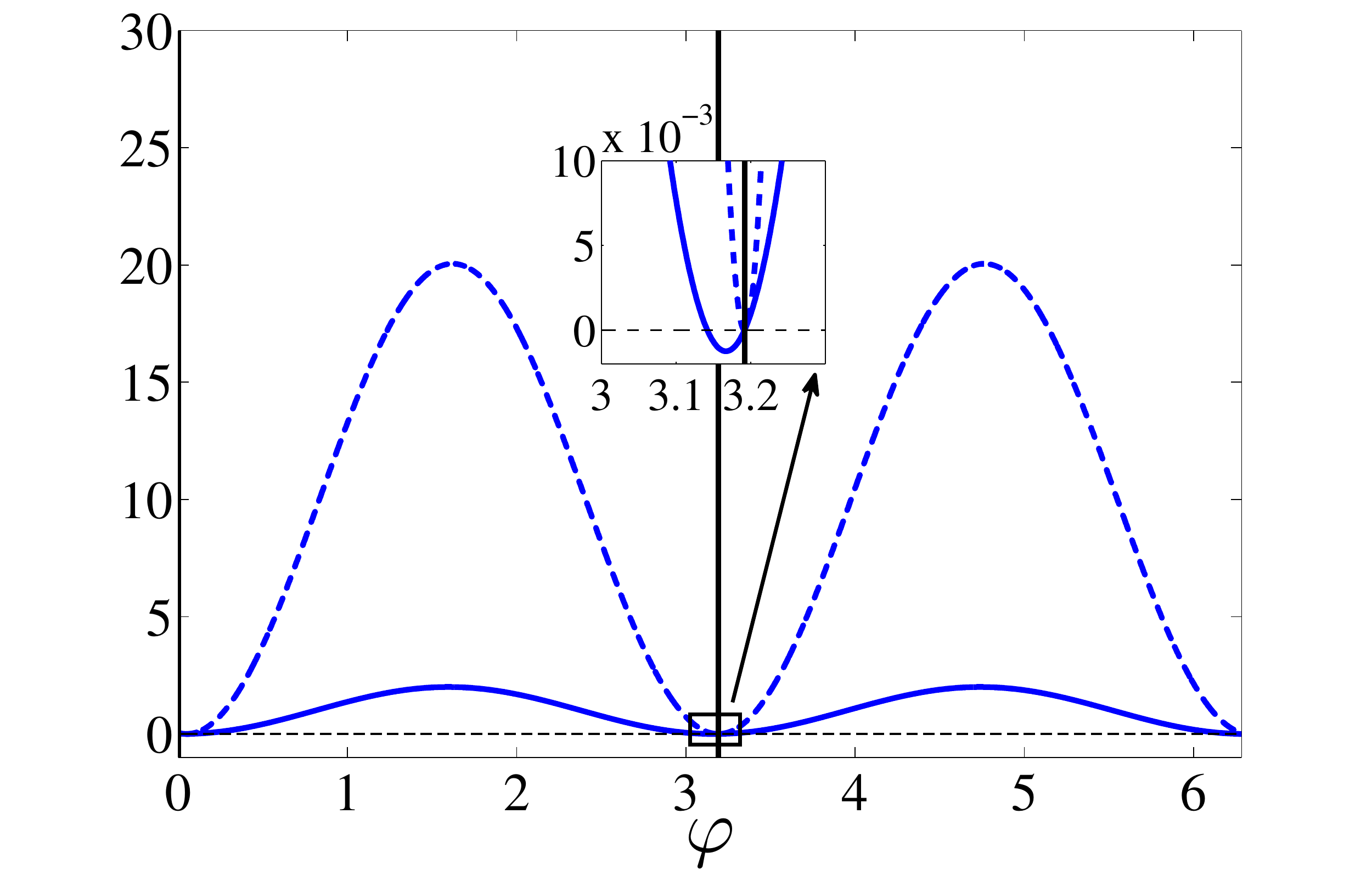}\includegraphics[scale=.2]{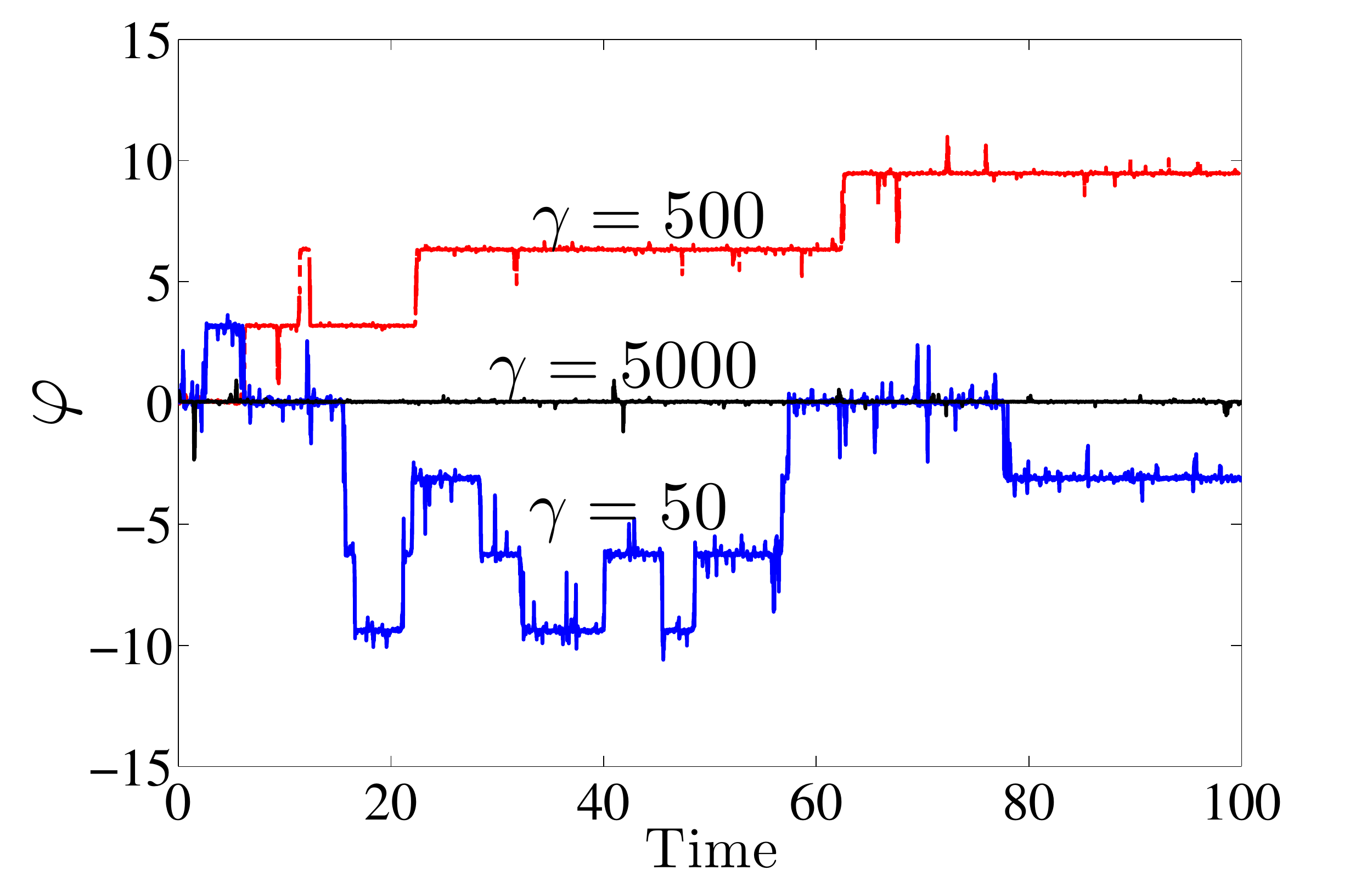}}\\
\subfigure[\hspace{9cm}]{\includegraphics[scale=.2]{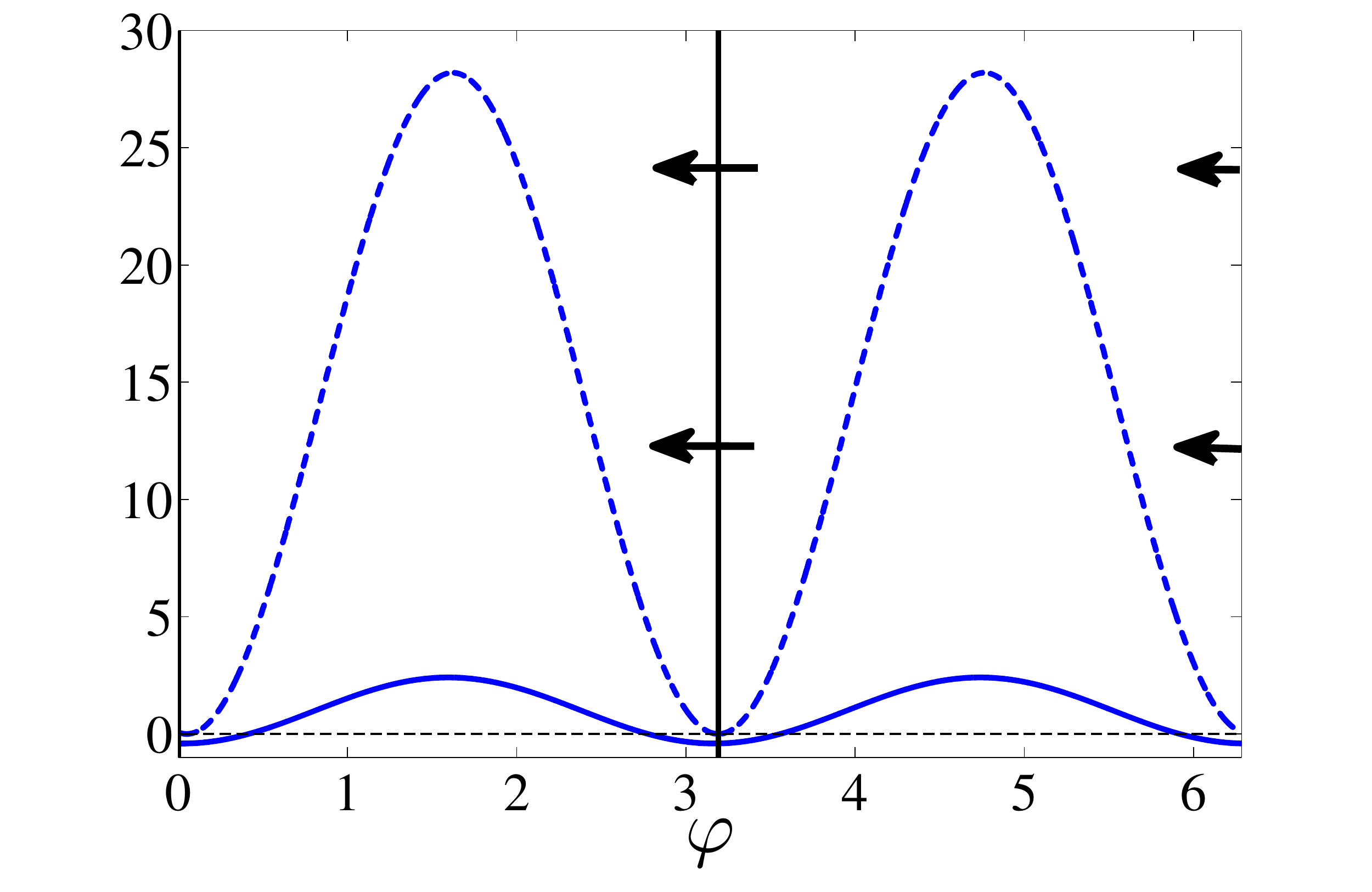}\includegraphics[scale=.2]{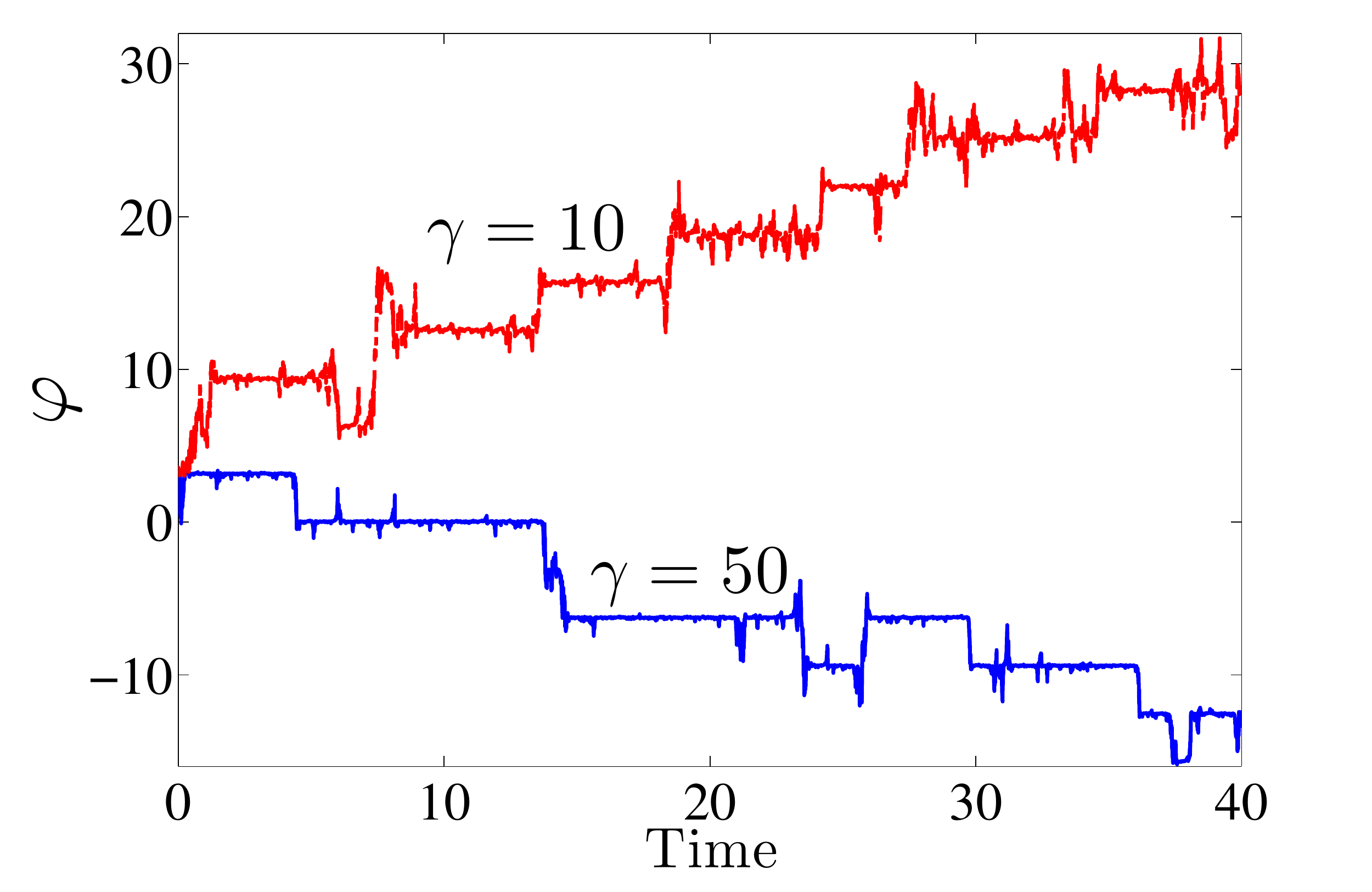}}
\end{center}
\caption{{\bf CR system is diffusion dominated.}  Left panels show the Ito drift (solid lines) and diffusivity (dashed lines) of the phase reduced system \eqref{eq:phase}.  Solid vertical lines indicate the ratchet points (where the diffusivity is zero) and the horizontal arrows indicate the sign of the Ito drift at the ratchet points.  Right panels show sample phase trajectories of the full system \eqref{eq:rphi}. For all plots, $c=20$, $\omega=1$, and unless indicated otherwise $\gamma=50$. (a) $K_{CR}=4$, (b) $K_{CR}=6.32$, and (c) $K_{CR}=7.5$.}\label{fig:CRItodiff}
\end{figure}

For the CR system, the ratchet points (zeros of the diffusivity) occur at approximately $\varphi^*=0,\pi$ as $D_{\varphi}^{CR}(\varphi^*)=\mathcal{O}(\varepsilon^2)$ (see \eqref{eq:CRexpansions}) independent of $K_{CR}$. The left panel Figure \ref{fig:CRItodiff} (a) plots the Ito drift and diffusivity for the CR system when $K_{CR}=4$.  In this case, the diffusivity is much larger than the Ito drift except near the two ratchet points. Thus, switching between the metastable states (zeros of the Fickian drift with negative slope) will occur at a higher rate than in the AR system. At the ratchet points, the Ito drift is positive, which indicates that the reduced phase system \eqref{eq:phase} predicts that the process moves past the ratchet points in the direction of the deterministic limit cycle.  A sample trajectory from the full system \eqref{eq:rphi} in the right panel shows that this is indeed the case as $\varphi$ increases with time.  

Figure \ref{fig:CRItodiff} (b) shows the behavior of the CR system when $K_{CR}=\sqrt{2\omega/\varepsilon} \approx 6.32$ where $I_{\varphi}^{CR}(\varphi^*)=\mathcal{O}(\varepsilon^2)$ (see \eqref{eq:CRexpansions}).  Thus, the Ito drift is approximately zero at the ratchet points.  As in the AR system, the phase reduction predicts that once a trajectory arrives at a ratchet point, it will be stuck there indefinitely.  However, the sample trajectories in the right panel again show that accuracy of this predicition is highly dependent on the value of $\gamma$.  Lastly, when $K_{AR}>\sqrt{2\omega/\varepsilon}$ (Fig. \ref{fig:ARItodiff} (c)), the Ito drift becomes negative at the ratchet points, and the phase reduction predicts that the process can only move past the ratchet points in the opposite direction of the deterministic limit cycle. In contrast to the AR system, the CR system has a diffusivity that is \emph{larger} (in magnitude) than the Ito drift away from the ratchet points.  Thus, the switching in the opposite direction on the limit cycle will occur at a higher rate than in the AR system (compare right panel of Fig \ref{fig:CRItodiff} (c) to right panel in \ref{fig:ARItodiff} (d)).  Just as in the AR system, if $\gamma$ is made smaller, the full system can also overcome the ratchet points in the same direction of the deterministic limit cycle by moving off the limit cycle (right panel, $\gamma=10$ trace).

\begin{figure}[h!]
\begin{center}
\includegraphics[scale=.35]{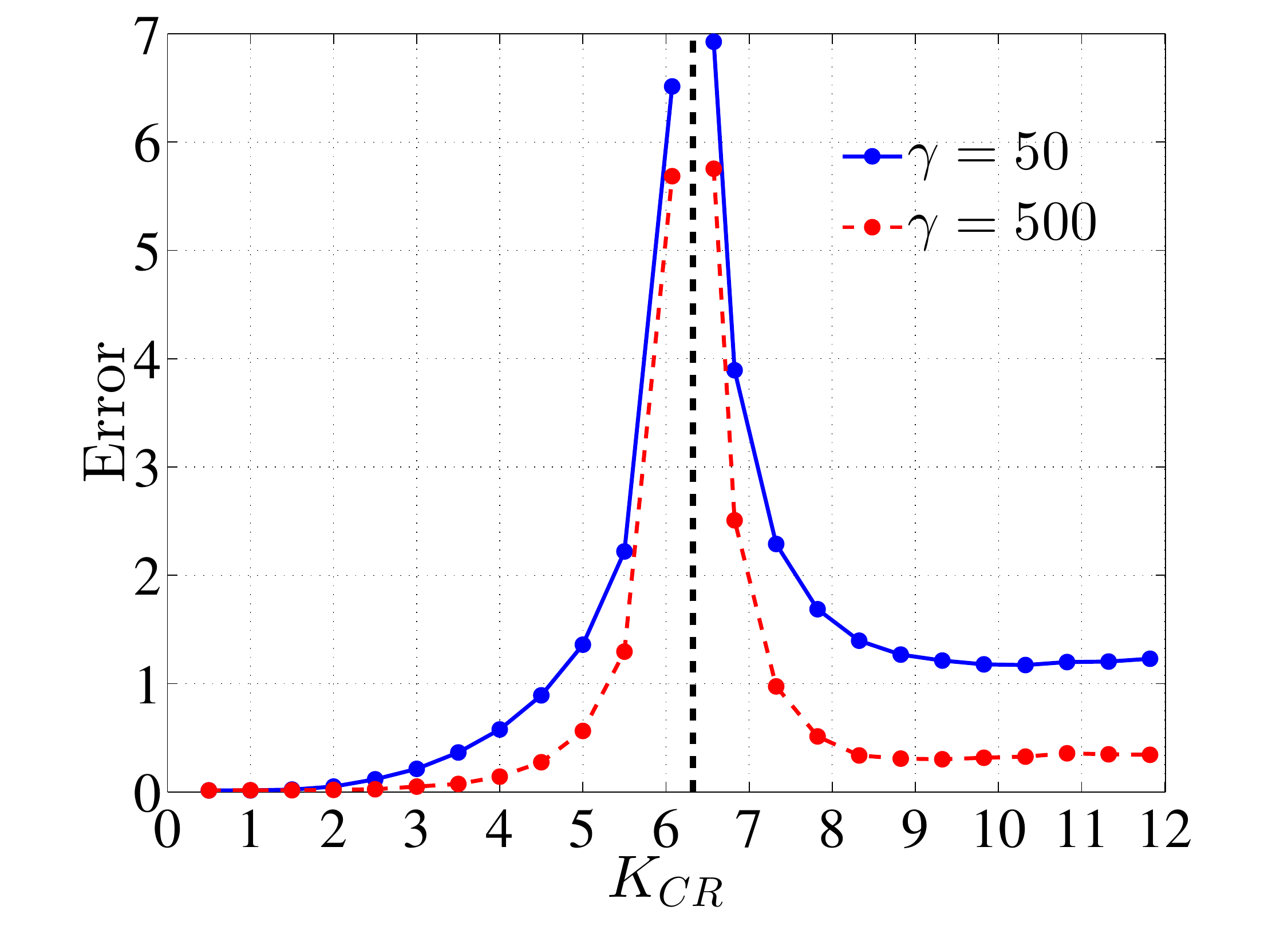}
\end{center}
\caption{{\bf Error in the phase model approximation of the stationary density for the CR system as a function of $K_{CR}$.} $L^2$ norm of the difference between the stationary density obtained from solving \eqref{eq:phaseFPIto} and the stationary density obtained from Monte-Carlo simulations of the full system \eqref{eq:rphi} as a function of $K_{CR}$ when $c=20$, $\omega=1$, $\gamma=50$ (solid line), and $\gamma=500$ (dashed line) }\label{fig:CRerror}
\end{figure}

Just as we did with the AR system, we quantify the accuracy of the phase reduced system \eqref{eq:phase} by exploring the error \eqref{eq:error} between the stationary density obtained from solving the FP equation for the reduced system \eqref{eq:phaseFPIto} $u_{ss}$ and the stationary density obtained from Monte-Carlo simulation $u_{ss}^{MC}$. Figure \ref{fig:CRerror} plots this error as a function of $K_{CR}$ and two different values for $\gamma$.  Interestingly, the error overall appears to be larger than what we found in the AR system (compare y-axis scale in Figures \ref{fig:CRerror} and \ref{fig:ARerror}).  Nonetheless, the qualitative trend is the same in both systems: as $K_{CR}$ approaches the value where the Ito drift is zero at the ratchet points, the error in the phase reduction increases. Furthermore, increasing $\gamma$ causes the error to decrease for all values $K_{CR}\neq \sqrt{2\omega/\varepsilon}$ (compare solid and dashed lines).

\section{Discussion}

We have shown that a limit cycle system with additive noise can display a wide range of bistable switching dynamics.  Using stochastic phase reduction methods \cite{teramae2009, yoshimura2008}, we are able to relate different aspects of this bistable switching behavior to terms appearing in the phase reduced model.  That is, the emergence of bistable switching in a limit cycle system occurs when the drift term of the conservation (Fickian) form of the Fokker-Planck equation for the phase reduced system becomes negative.  The zeros of the Fickian drift with negative slope represent \emph{metastable} points that the stochastic process switches between and correspond to peaks in the stationary density for the phase.  The rate and directionality of the switching are controlled by the drift term in the stochastic differential equation for the phase reduced system, i.e., the Ito drift. We have shown that the Ito drift determines the stationary probability flux, and thereby the directionality of the switching.  That is, if the flux is positive (negative), the system will (on average) switch between the metastable points in the same (opposite) direction of the deterministic limit cycle.  Lastly, as with the Kramer's problem, the magnitude of the Ito drift relative to the diffusivity determines the rate at which the process switches between the metastable points. 

We illustrated these findings with the use of simple planar limit cycle systems that have areas in their vector fields where rotation occurs in the opposite direction of the limit cycle.  In the CR system, this counterrotation occurs on only one side of the limit cycle.  This lead to the diffusivity the phase reduced system being much larger in magnitude than the Ito drift.  Thus, the CR system displays fairly frequent swithching between metastable points.  In the AR system, the counterrotation occurs both inside and outside of the limit cycle.  The phase reduction for the AR system had a diffusivity that was much smaller in magnitude than the Ito drift, and thus switching between metastable points is a rare event.  Both systems can display switching in either the same or opposite direction of the deterministic limit cycle.  Thus, it is the differences in the dynamics off the limit cycle between the two systems that is responsible for the different rates of switching. This highlights the fact that in stochastic limit cycle systems, if the interactions of the noise with the vector field off the limit cycle are ignored in the phase reduction (i.e., the $\mathcal{O}(\sigma^2)$ terms in \eqref{eq:phase}), then the phase model will fail to capture many interesting phenomena that are present in the full system \eqref{eq:XY}.   

We have also identified an interesting singularity where the leading order phase reduced system fails to capture the dynamics of the full system regardless of the rate of attraction back to the limit cycle. This occurs when the Ito drift of the phase reduced system has a zero at exactly the same point as the diffusivity.  In this case, the phase reduction predicts that the system should remain stuck at this point for all time.  However, simulations of the full system reveal that this is not the case. Thus, $\mathcal{O}(1/\gamma)$ terms that are ignored in the leading order reduction become necessary to capture the behavior of the full system.  

This work identifies an alternative mechanism for the emergence of bistable switching dynamics observed in a number of physical systems ranging from the dynamics of genetic swiches in bacteria \cite{lai2004,gardner2000,veening2008} to up and down states in the cortex thought to be involved in working memory \cite{cossart2003,fuster1995}.  Interestingly, the dynamics of the phase variable in the CR system (Figure \ref{fig:bistablephase} (a)) is reminiscent of the behavior of a brownian ratchet, which is a standard model for molecular motors \cite{astumian1994,kolomeisky2007,svoboda1993}.  Furthermore, as this bistable switching behavior has also been shown to occur in more complex limit cycle systems \cite{newby2014}, this could lead to interesting consequences for coupled oscillator systems.  Indeed, an interesting type of mixed synchronization has already been observed when oppositely rotating oscillators are coupled together (see, for example, \cite{bhowmicketal2011,prasad2010}).  Exploring the behavior of coupled bistable oscillators could lead to new synchronization or phase-locking phenomena.


\section{Acknowledgements}

The authors are supported in part by the Mathematical Biosciences Institute and the National Science Foundation under Grant No. DMS-0931642.



\end{document}